\documentclass[conference]{IEEEtran}
\IEEEoverridecommandlockouts

\usepackage{cite}
\usepackage{tikz}
\usepackage{amsmath,amssymb,amsfonts}
\usepackage{algorithmic}
\usepackage{graphicx}
\usepackage{textcomp}
\usepackage{filecontents}
\usepackage{xspace}
\usepackage{comment}
\usepackage[hidelinks]{hyperref}
\usepackage{balance}
\usepackage[capitalize,noabbrev]{cleveref}
\crefformat{section}{\S#2#1#3}

\usepackage{multirow}
\usepackage{color}
\usepackage{subcaption}
\usepackage{caption}
\usepackage{makecell}

\usepackage{listings}
\usepackage{xcolor}
\usepackage{ragged2e}

\usepackage[most]{tcolorbox}
\usepackage{enumitem}
\usepackage{tabularx}
\usepackage{booktabs} 

\definecolor{codegreen}{rgb}{0,0.6,0}
\definecolor{codegray}{rgb}{0.5,0.5,0.5}
\definecolor{codepurple}{rgb}{0.58,0,0.82}
\definecolor{backcolour}{rgb}{0.95,0.95,0.92}
\definecolor{aliceblue}{rgb}{0.95, 0.975, 1.0}
\definecolor{special}{rgb}{0.97, 0.97, 0.97}

\lstdefinestyle{mystyle}{
    backgroundcolor=\color{special},   
    commentstyle=\color{codegreen},
    keywordstyle=\color{black},
    numberstyle=\tiny\color{codegray},
    stringstyle=\color{codepurple},
    basicstyle=\small\ttfamily\footnotesize,
    breakatwhitespace=true,         
    breaklines=true,                 
    captionpos=b,                    
    keepspaces=true,                                  
    showspaces=false,                
    showstringspaces=false,
    showtabs=false,
    breakautoindent=false,
    breakindent=0ex,
    tabsize=2,
    frame=single,
    columns=fullflexible, 
    keepspaces=true
}

\newcommand{\paraspace}{\vspace{0.05in}} 

\newcommand{\parab}[1]{\paraspace\noindent{\bf #1}}

\newcommand{\eg}{{\it e.g.},~}
\newcommand{\etal}{{\it et al.~}}

\newcommand{\ie}{{\it i.e.},~}

\newenvironment{icompact}{%
  \begin{list}{$\bullet$}{%
    \setlength{\leftmargin}{1.2em}    
    \setlength{\itemindent}{0pt}%
    \setlength{\labelwidth}{0pt}%
    \setlength{\labelsep}{0pt}%
    \setlength{\parsep}{0.5pt plus 0.5pt}%
    \setlength{\partopsep}{0.5pt plus 0.5pt}%
    \setlength{\topsep}{0.5pt plus 1pt minus 0.5pt}%
    \setlength{\itemsep}{0.5pt plus 0.5pt}%
    \setlength{\parskip}{0pt plus 1pt}%
  }%
}{%
  \normalsize\end{list}
}

\def\BibTeX{{\rm B\kern-.05em{\sc i\kern-.025em b}\kern-.08em
    T\kern-.1667em\lower.7ex\hbox{E}\kern-.125emX}}

\def\sys{\textsc{RouteLLM}\xspace}

\IEEEoverridecommandlockouts\IEEEpubid{\makebox[\columnwidth]{$\copyright$ 2026 IEEE. Personal use of this material is permitted.\hfill}\hspace{\columnsep}\makebox[\columnwidth]{ }}

\begin{document}
\date{}

\title{\Large \bf The Surprising Effectiveness of LLMs in BGP Security: Mining An Unprecedented Amount of Incidents and Boosting Anomaly Detection}

\author{
\IEEEauthorblockN{
Libin Liu$^{\S}$,
Wenzhou Yang$^{\dagger}$, 
Li Chen$^{\S}$, 
Dan Li$^{\ddagger}$, 
Xiuting Xu$^{\S}$
}
\IEEEauthorblockA{
$^{\S}$Zhongguancun Laboratory,
$^{\dagger}$Beijing University of Posts and Telecommunications, 
$^{\ddagger}$Tsinghua University
}
}

\maketitle

\begin{abstract}
Border Gateway Protocol (BGP) security is critical to Internet infrastructure, yet progress in routing anomaly detection has been limited by the scarcity of publicly available incident datasets, which contain only 18 recorded cases. 
We observe that public operator mailing lists, \eg NANOG and AusNOG, contain abundant yet largely untapped reports of real-world routing anomalies. 
To leverage this source, we develop an LLM-assisted extraction pipeline that identifies 244 candidate incidents from historical discussion threads. After expert validation, we curate a verified benchmark containing 232 confirmed routing anomaly events, making it 11.89$\times$ larger than existing dataset.

Using this benchmark, we show that existing routing anomaly detection systems generalize poorly to diverse real-world incidents. 
At the same time, we find that some general-purpose LLMs without routing-specific adaptation can identify a subset of routing anomalies, but their performance varies across models and remains insufficient for reliable routing anomaly detection.
Motivated by this observation, we design \sys, an LLM-based routing anomaly detector that integrates BGP-semantic-aware tokenization, routing-domain adaptation, and time-aware routing evidence retrieval.
Experimental results show that \sys achieves 87.13\% event-level accuracy and 94.65\% message-level accuracy, outperforming the strongest baselines by 55.30\% and 68.50\%, respectively.
We open-source the verified routing anomaly benchmark, fine-tuned model, and implementation code to support future research on BGP security\footnote{https://github.com/vvcode1/RouteLLM}.
\end{abstract}

\begin{IEEEkeywords}
BGP Security, Routing Anomaly Incidents, Routing Anomaly Detection, Large Language Model
\end{IEEEkeywords}

\section{Introduction}\label{sec:intro}

Securing inter-domain routing remains a critical challenge for the global Internet. Despite decades of research and the proposal of numerous security enhancements, such as BGPsec~\cite{bgpsec2017}, BGPiSec~\cite{bgpisec2024}, psBGP~\cite{psbgp2007}, and S-BGP~\cite{sbgp2000}, deployment remains limited due to architectural incompatibilities and operational overhead. Meanwhile, the adoption of light-weight mechanisms such as ROA~\cite{roa2013} and ASPA~\cite{aspa2025} is ongoing, but their effectiveness is constrained by partial deployment and the lack of Route Origin Validation (ROV)~\cite{chung2019rpki}.
As a result, detecting anomalous BGP behavior remains an indispensable complementary defense~\cite{beam2024,bgpmon_route_monitor,cloudflare_radar}.

A fundamental obstacle to advancing BGP anomaly detection is the lack of large-scale, real-world routing anomaly incident datasets. 
Existing systems are all evaluated using a very small number of incidents, at most 18 anomalies over the past two decades ($\S$\ref{sec:missing_ra_data}), or rely on simulated attacks. This data sparsity restricts the generalizability and robustness of detection systems, especially as routing anomalies evolve in sophistication and scale~\cite{goldberg2014taking,VPAT2015}.

We observe that public network operator mailing lists, such as NANOG~\cite{nanog2025}, AfNOG~\cite{afnog2025}, and RIPE NOGs~\cite{ripenogs2025}, contain rich untapped information on real-world routing anomalies (\cref{sec:mail_lists_intro}). Operators frequently report incidents, share logs, and coordinate mitigations via these channels. These reports often include sufficient detail, such as AS numbers, affected prefixes, and routing behaviors, to label ground-truth events. However, mining these sources is nontrivial: the text is unstructured, jargon-heavy, and mixed with unrelated operational discussions. For instance, NANOG alone contains over 286,292 emails authored by more than 9,500 individuals.

Recent advances in large language models (LLMs)~\cite{AEFJ2025,ZWDY2024,docetl2024,wu2024netllm} provide a potential solution. LLMs excel at processing unstructured text and extracting structured knowledge. We investigate whether LLMs can be used to automatically extract routing anomaly events from operator mailing lists. We prompt LLMs to parse entire email threads, identify anomaly reports, and extract structured incident data. 
We apply this approach to thirteen mailing lists covering more than three decades of discussions, yielding 244 candidate incidents (\cref{sec:data_collection}).
To ensure reliability, we cross-validate results across multiple LLMs (GPT-5~\cite{gpt-5}, Gemini-3-Pro~\cite{gen-3}, and Claude-4.5-Haiku~\cite{haiku}) and verify final outputs with human experts. In the end, we obtain a curated dataset of 232 distinct anomaly events (\cref{sec:expert_curation}),
which is 11.89$\times$ larger than the existing one.

We then revisit state-of-the-art (SOTA) routing anomaly detection systems using this expanded dataset. We evaluate BEAM~\cite{beam2024} 
on the newly collected events and find that its detection accuracy degrades significantly: BEAM achieves only 57.67\% event-level accuracy (\cref{sec:soda_perf}). 
This gap suggests that existing systems may be overfitted to a narrow set of previously known anomalies (\cref{sec:motivate_routellm}).

Motivated by the success of LLMs in related networking and security tasks~\cite{HPSP2024,MRMM2024,FCMN2024,HJVm2023,guthula2023netfound}, we ask a natural question: \textit{Can LLMs improve routing anomaly detection?} To answer this, we evaluate general-purpose LLMs on the same detection task, without any domain-specific adaptation. 
Models like Llama-3.1-8B-Instruct achieve 54.24\% accuracy on event-level anomaly detection, only 5.95\% below BEAM~\cite{beam2024}, the state-of-the-art system (\cref{sec:llm_perf}). However, the performance varies across models and remains insufficient for reliable routing anomaly detection.

Through manual inspection of representative false positives and false negatives, we find that LLMs often treat BGP updates as ordinary text, fail to capture protocol-level routing semantics, and make predictions without grounding on time-consistent routing evidence such as prefix ownership and AS relationships.
These observations lead to three key insights for improving LLM-based routing anomaly detection:
(1) LLMs should understand the structured semantics of BGP protocol data rather than flattening routing fields into plain text;
(2) LLMs should acquire routing-domain knowledge, such as AS relationships and routing anomaly patterns;
and (3) routing decisions should be grounded on time-consistent external routing evidence because Internet routing information continuously evolves (\cref{sec:motivate_routellm}).

Guided by these insights, we develop \sys, a routing-domain-adapted LLM framework for routing anomaly detection.
\sys improves the routing anomaly detection capability of LLMs through three key designs.
First, 
to enable LLMs to better understand structured BGP protocol data, \sys introduces a BGP-semantic-aware tokenizer that treats routing constructs, such as AS paths and announcements, as atomic semantic units (\cref{sec:token_design}).
Second, to inject routing-domain knowledge into the model, \sys performs dual-stage fine-tuning on AS relationship data and labeled BGP anomaly datasets using parameter-efficient LoRA adaptation~\cite{lora2021} (\cref{sec:fine_tune_design}).
Although incident-level labels are limited, each incident contains many associated BGP updates and routing-evidence records, which are sufficient for adapting pretrained LLMs to routing anomaly detection~\cite{guthula2023netfound,wu2024netllm}.
Third, to account for the dynamic nature of Internet routing, \sys incorporates a time-aware retrieval-augmented generation (RAG) pipeline that provides timestamp-consistent routing evidence, including AS relationships and prefix attribution information (\cref{sec:rag_design}).

In this work, we adopt Llama-3.1-8B-Instruct as the base model and apply the \sys design pipeline to fine-tune it for routing anomaly detection.
We choose this model because it provides a strong balance between reasoning capability, deployment cost, and fine-tuning efficiency for routing-domain adaptation (\cref{sec:llm_perf}).
Importantly, the \sys pipeline is model-agnostic and can be extended to other LLMs to further improve their routing anomaly detection capability.

In summary, we make the following contributions.
\begin{itemize}
    \item We highlight the scarcity of existing routing anomaly incidents and demonstrate that operator mailing lists can be mined to extract high-quality anomaly incidents (\cref{sec:bg}). 
    \item We develop an automatic routing anomaly incident collection pipeline with LLMs. With specially designed prompts and expert curation, we collect 11.89$\times$ more incidents than prior work (\cref{sec:dataset}).  
    \item We use the collected routing incident dataset to evaluate a representative detector and vanilla LLMs, revealing a generalization gap in existing systems and the limitations of general-purpose LLMs without routing-domain adaptation (\cref{sec:preliminary}).
    \item We design and implement \sys, the first LLM-based routing anomaly detector. It bridges the gap between BGP data and LLMs, integrates routing knowledge through fine-tuning, and incorporates time-consistent external routing evidence through a dedicated retrieval pipeline (\cref{sec:design}, \cref{sec:sys_imple}).
    \item We evaluate \sys through testbed experiments. \sys achieves 87.13\% average event-level accuracy and 94.65\% average message-level accuracy, outperforming the compared schemes by at least 55.30\% and 68.50\%, respectively (\cref{sec:eval}).
\end{itemize}
\section{Background and Motivation}\label{sec:bg}

In this section, we first discuss the scarcity of routing anomaly data in prior research (\cref{sec:missing_ra_data}) and explain why public anomaly detection systems cannot directly provide ground-truth incident labels (\cref{sec:pub_data_sys}). We then explore the potential of public mailing lists as a complementary data source (\cref{sec:mail_lists_intro}).

\subsection{Scarcity of Routing Anomaly Data}\label{sec:missing_ra_data}

\begin{table*}[t]
\centering
\caption{The routing anomaly incident data used in existing literature.}
\vspace{-2mm}
\label{tab:stat_ab_events}
\scalebox{1.05}{
\begin{tabular}{c||cccc|c}
\hline\hline
\multirow{2}{*}{Existing Work}         & \multicolumn{4}{c|}{Number of Real-world Routing Anomaly Incidents}                                                                                                          & \multirow{2}{*}{Total Number} \\ \cline{2-5}
                                       & \multicolumn{1}{c|}{Exact Prefix Hijacking} & \multicolumn{1}{c|}{Sub-prefix Hijacking} & \multicolumn{1}{c|}{Path Hijacking} & \multicolumn{1}{c|}{Route Leak} &                                                       \\ \hline
Zheng. \etal\cite{ZCJL2007}            & \multicolumn{4}{c|}{Synthetic data}   & -                                                                                                                                        \\ \hline
Xin. \etal\cite{HXMZ2007}              & \multicolumn{4}{c|}{No ground-truth data}  & -                                                                                                                                       \\ \hline
iSPY~\cite{ispy2008}                   & \multicolumn{4}{c|}{Synthetic data}    & -                                                                                                                                    \\ \hline
Buddyguard~\cite{buddyguard2012}       & \multicolumn{1}{c|}{2}                      & \multicolumn{1}{c|}{1}                    & \multicolumn{1}{c|}{-}              & \multicolumn{1}{c|}{1}          & 4                                                     \\ \hline
Pierre. \etal\cite{VPAT2015}           & \multicolumn{1}{c|}{7}                      & \multicolumn{1}{c|}{-}                    & \multicolumn{1}{c|}{3}              & \multicolumn{1}{c|}{-}          & 10                                                    \\ \hline
MS-LSTM~\cite{mslstm2016}             & \multicolumn{4}{c|}{Not specifying the anomaly types}   & 6                                                                                                                                                  \\ \hline
Artemis~\cite{artemis2018}             & \multicolumn{4}{c|}{Synthetic data}    & -                                                                                                                                      \\ \hline
Cecilia. \etal\cite{TCRP2019}           & \multicolumn{4}{c|}{Not specifying the anomaly types} & 3                                                                                                                                                   \\ \hline
Odnan. \etal\cite{SOFS2019}             & \multicolumn{4}{c|}{Not specifying the anomaly types} & 4                                                                                                                                                    \\ \hline
Yutao. \etal\cite{DYLQ2021}             & \multicolumn{4}{c|}{Results of other anomaly detection systems}  & -                                                                                                     \\ \hline
DFOH~\cite{HTAT2024}                   & \multicolumn{1}{c|}{-}                      & \multicolumn{1}{c|}{2}                    & \multicolumn{1}{c|}{-}              & \multicolumn{1}{c|}{-}          & 2                                                     \\ \hline
BEAM~\cite{beam2024}                 & \multicolumn{1}{c|}{2}                      & \multicolumn{1}{c|}{13}                   & \multicolumn{1}{c|}{-}              & \multicolumn{1}{c|}{3}          & 18                                                    \\ \hline
BGPiSec~\cite{bgpisec2024}           & \multicolumn{4}{c|}{Synthetic data}    & -                                                                                                                                                                       \\ \hline
Ares~\cite{taoares}         & \multicolumn{1}{c|}{-}                      & \multicolumn{1}{c|}{-}                    & \multicolumn{1}{c|}{12}              & \multicolumn{1}{c|}{-}          & 12                                                      \\ \hline\hline 
\end{tabular}
}
\vspace{-3mm}
\end{table*}

Despite the critical role of routing anomaly detection in enhancing Internet security, 
the research community still lacks large-scale, real-world ground-truth routing anomaly data.
Prior studies have predominantly relied on a limited set of manually verified incidents or synthetic data through simulation. 
In this study, we review major research efforts in routing anomaly detection from the past two decades and analyze the data used in their evaluations, as summarized in Table~\ref{tab:stat_ab_events}.

Among the surveyed literature, four studies~\cite{ZCJL2007,ispy2008,artemis2018,bgpisec2024} explicitly report the use of synthetic data.
Xin \etal\cite{HXMZ2007} acknowledge employing BGP data lacking ground truth, thus depending on confirmation from affected networks to support their findings.
Yutao \etal\cite{DYLQ2021} utilize outputs from other detection systems as reference labels and admit that these data may not be reliable.
MS-LSTM~\cite{mslstm2016}, Cecilia \etal \cite{TCRP2019}, and Odnan \etal\cite{SOFS2019} use six, three, and four anomaly incidents, respectively, though without specifying anomaly types.
Although Buddyguard~\cite{buddyguard2012}, Pierre \etal\cite{VPAT2015}, DFOH~\cite{HTAT2024}, BEAM~\cite{beam2024}, and Ares~\cite{taoares} use more anomalies, up to 18 events, none of them cover the full range of anomaly categories, \ie prefix hijacking, path hijacking, and route leak.

Access to up-to-date and large-scale real-world routing anomaly data is crucial for the development of effective detection systems.
The lack of such data limits progress in both accuracy and generalizability. 
Statistical methods~\cite{ZCJL2007,VPAT2015,heap2016,HXMZ2007,buddyguard2012,argus2012,LJDD2005,TCRP2019,CSFR2019} rely on a comprehensive analysis of normal and anomalous routing data to define detection rules. 
Meanwhile, advanced machine learning  approaches~\cite{mslstm2016,DYLQ2021,rqa2015,ARNM2012,LABM2014,STSY2020,HKTP2021,ap2vec2022,MKAH2019,HTAT2024} require large-scale datasets for training.
However, dataset scarcity remains a substantial obstacle, 
and synthetic data often fail to capture the true complexity of real-world network environments~\cite{HTAT2024}.

\subsection{Public Alerts Are Not Ground Truth}
\label{sec:pub_data_sys}

Public anomaly detection platforms, such as Cloudflare Radar~\cite{cloudflare_radar} and the former BGPmon service~\cite{bgpmon_route_monitor}, provide valuable real-time visibility into potential BGP anomalies. 
However, while these services offer broad coverage and high-frequency updates, their data suffer from critical limitations that render it unsuitable for directly training and evaluating advanced anomaly detection systems.

First, the alerts generated by these systems are algorithmically triggered inferences rather than confirmed routing incidents. 
They are based on predefined heuristic thresholds and pattern matching techniques that prioritize rapid detection over accuracy~\cite{cloudflare,cloudflare-leak_dec}. 
For example, Cloudflare Radar~\cite{cloudflare_radar} outputs routing anomalies with an associated confidence score, reflecting the heuristic nature of their methodology rather than representing verified network events~\cite{cloudflare}. 
Consequently, these data feeds exhibit a high false positive rate, incorporating a substantial share of events that are later deemed benign, such as routine maintenance or traffic engineering, rather than malicious routing behavior. 
Also, through discussions with network operators at multiple ISPs that employ similar internal monitoring tools, 
we confirm that the overwhelming majority of such automated alerts require subsequent manual review and are frequently classified as non-actionable. 
This inherent noise introduces label inaccuracy that misleads the training and evaluation of detection systems.

Moreover, these systems cannot provide ground-truth
labels~\cite{cloudflare_radar,bgpmon_route_monitor}: their alerts are unverified, making true positives
indistinguishable from false alarms without additional manual analysis.
Consequently, prior work avoids using such alerts directly as benchmark
or training labels~\cite{mslstm2016,rqa2015,ARNM2012,LABM2014,STSY2020,HKTP2021,ap2vec2022,MKAH2019,HTAT2024}, since they lack the verification required
for reliable model development.

\subsection{Anomaly Incident Reports from Public Mailing Lists}\label{sec:mail_lists_intro}

\begin{figure}[t]
\centering
\begin{minipage}{0.45\textwidth}
\begin{lstlisting}
(*\bfseries Subject*): BGP route hijack by AS10990
(*\bfseries From*): Clinton Work clinton at scripty.com
(*\bfseries Time*): (*\textit{Thu Jul 30 02:46:55 UTC 2020}*)
(*\bfseries Contents*): We saw a bunch of our IP blocks hijacked by AS10990 from 19:15 MDT until 20:23 MDT. Anybody else have problems with that.

ASpath:  1299 7219 10990

50.92.0.0/17	AS10990
198.166.0.0/17	 AS10990
198.166.128.0/17	AS10990
162.157.128.0/17	AS10990
162.157.0.0/17	AS10990
50.92.128.0/17	AS10990

--
Clinton Work
Airdrie, AB
\end{lstlisting}
\end{minipage}
\vspace{-3mm}
\caption{An example of prefix hijacking incident report from the NANOG mailing list~\cite{hijackexp}.}
\vspace{-5mm}
\label{fig:exp_abnormal_report}
\end{figure}

In real-world network operations, 
AS-level network operators maintain filter lists, transit policies, and other routing configurations~\cite{clemens2021revisiting}.
They actively monitor network conditions and share alerts about anomalies such as network failures and suspicious BGP announcements. 
Rather than relying solely on automated tools, operators often use public mailing lists as both real-time communication platforms and historical archives.
In addition, unlike data obtained directly from anomaly detection systems, the incidents reported on these mailing lists are contributed by operators of the ASes that own the affected IP resources. 
As a result, 
these mailing lists can serve as a reliable source of validated information for tracking real-world routing anomalies.

Fig.~\ref{fig:exp_abnormal_report} shows an example of prefix hijacking report from NANOG mailing list.
An operator reports that some IP prefixes are hijacked by AS10990 and announced along the AS path ``1299 7219 10990'', and also states the time when the incident is first observed and its duration time. 
From this, we can extract key details of the hijacking event, including victim prefixes, misbehaving AS (AS10990), occurrence time (2020-07-30 01:15 UTC), and its duration time ($\sim$68 minutes).
Although the victim network is not explicitly stated in this report, we can identify it based on the operator's organization information and public databases, such as WHOIS~\cite{whois} and PeeringDB~\cite{pdb}.
\section{Building a Verified Routing Anomaly Benchmark}\label{sec:dataset}

\begin{figure}[t]
  \centering
    \includegraphics[width=1.00\columnwidth]{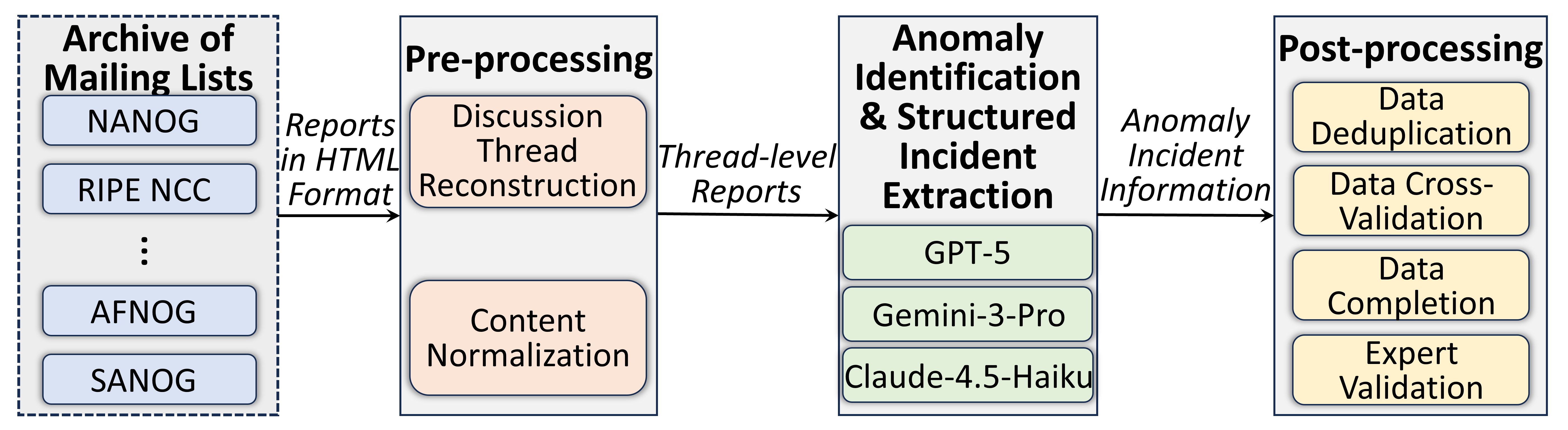}
    \vspace{-5mm}
    \caption{The workflow for building a verified routing anomaly benchmark.}
    \vspace{-2mm}
  \label{fig:data_collection_process}
\end{figure}

In this section, we describe how we construct a verified benchmark of real-world routing anomalies from public mailing lists, as shown in Fig.~\ref{fig:data_collection_process}.
Our goal is not merely to collect more incidents, but to build a benchmark whose labels can be traced back to operator-provided evidence.
We first describe the data sources and thread reconstruction process (\cref{sec:data_sources}), then present our LLM-based evidence-aware extraction pipeline (\cref{sec:data_collection}), and expert validation procedure (\cref{sec:expert_curation}).

\subsection{Data Sources and Thread Reconstruction}\label{sec:data_sources}

We collect public operator discussions from thirteen mailing lists, including NANOG~\cite{nanog2025}, RIPE NCC~\cite{ripenogs2025}, SANOG~\cite{safnog2025}, AusNOG~\cite{ausnog2025}, AfNOG~\cite{afnog2025}, INNOG~\cite{innog2025}, PACNOG~\cite{pacnog2025}, LACNOG~\cite{lacnog2025}, ITNOG~\cite{itnog2025}, SAFNOG~\cite{safnog2025}, SGNOG~\cite{sgnog2025}, NZNOG~\cite{nznog2025}, and CaribNOG~\cite{caribnog2025}. These lists are widely used by network operators around the globe and provide long-running public archives where operators report routing anomalies, share BGP paths, discuss reachability problems, and confirm mitigation outcomes. 

Mailing-list archives contain individual emails rather than self-contained incident reports. However, routing anomalies are often discussed across multiple replies: an initial message may report a suspicious route, while later replies may identify the affected prefix, clarify the responsible AS, or confirm that the problem has been resolved. 
Therefore, we reconstruct discussion threads before incident extraction. We group emails using message identifiers, reply headers, normalized subjects, and temporal proximity. For each thread, we preserve the original message order and retain quoted or forwarded content.

\subsection{Routing Anomaly Incident Identification and Extraction}\label{sec:data_collection}

We use LLMs as semantic analysis tools to extract structured routing anomaly incidents from operator discussions,
leveraging their demonstrated human-like comprehension capabilities and broad technical knowledge~\cite{chatgpt}.
Public mailing-list text is highly unstructured and often mixes natural language, quoted replies, raw BGP paths, traceroutes, and router outputs. 

As shown in Fig.~\ref{fig:data_collection_process}, our extraction pipeline consists of three main steps: content normalization, anomaly identification and structured incident extraction, and post-processing.
The pipeline is designed to be extensible, allowing for future incorporation of additional mailing lists as needed. 

\parab{Content normalization.}
We convert HTML emails into plain text and retain operational artifacts embedded in the discussion. We remove duplicated boilerplate and mailing-list footers, but preserve quoted messages because they may contain information that is necessary to understand the incident. 
We aggregate all emails in the same discussion thread as input for the next step.
This design is important because many routing anomaly reports are clarified through follow-up replies rather than fully described in the initial email.

\parab{Anomaly identification and structured incident extraction.}
For each thread, we provide the normalized discussion text to the LLM and ask it to determine whether the thread describes a routing anomaly. The model is instructed to consider prefix hijacks, path hijacks, and route leaks, while distinguishing them from unrelated outages, routine maintenance, traffic engineering, and general operational discussions.
If the thread is identified as a potential routing anomaly, we prompt the LLM to further extract a structured incident record, including the \textit{anomaly type}, \textit{attacker or leaking ASN}, \textit{victim ASN}, \textit{affected prefixes}, \textit{start and end timestamps}, and \textit{supporting textual evidence}. When a thread discusses multiple distinct incidents, the model is instructed to emit one record for each incident. We enforce a template-based JSON output format and apply schema validation to reject malformed records.

\parab{Post-processing.} After extraction, we integrate and normalize the candidate records. Specifically, we standardize timestamps to coordinated universal time (using email timestamps as fallbacks when precise event times are unavailable), normalize prefix lists into semicolon-delimited strings, and deduplicate redundant events based on anomaly type, involved ASes, affected prefixes, overlapping event times, and BGP-level evidence.
To improve extraction reliability, we cross-check the outputs from three LLM variants, including GPT-5, Gemini-3-Pro, and Claude-4.5-Haiku, and apply regex-based validation to all extracted fields. Fields with inconsistent outputs across models are flagged for expert review. This process yields 244 candidate routing anomaly incidents for subsequent validation.

\begin{table*}[t]
\centering
\caption{The routing anomaly detection accuracy of BEAM and general-purpose LLMs in terms of event-level and BGP message-level using the constructed benchmark. Here ``DS-R1-Qwen-7B'' represents the Qwen-7B model distilled by DeepSeek-R1, ``DS-V3'' represents DeepSeek-V3, and ``DS-R1'' represents DeepSeek-R1.}
\vspace{-2mm}
\label{tab:bg_models_acc}
\scalebox{0.95}{
\begin{tabular}{c||c|c|c|c|c|c|c|c|c}
\hline\hline
Accuracy      & BEAM    & DS-R1-Qwen-7B & Llama-3.1-8B & QwQ-32B & DS-V3 & DS-R1    & GPT-5  & Gemini-3-Pro   & Claude-4.5-Haiku \\ \hline 
Event-level   & 0.5767 & 0.5345   &   0.5424 & 0.3846  & 0.3607      & 0.5574       &  0.3904  &  0.3942  & 0.4255                \\ \hline
Message-level & 0.3671 & 0.3403   &   0.2430 & 0.1652  & 0.3134      & 0.1187       &  0.3578  &  0.4261  & 0.2399                 \\ \hline \hline
\end{tabular}
}
\vspace{-4mm}
\end{table*}

\subsection{Expert Validation}\label{sec:expert_curation}

To ensure the accuracy and operational relevance of the routing anomaly events extracted by the LLMs, we therefore introduce an expert validation phase to verify each candidate incident. Two domain experts with BGP and routing security expertise independently review each candidate incident. For each event, they examine the extracted record, the supporting text spans, and the full discussion thread. Rather than determining from raw BGP data whether an anomaly occurred, they verify that the discussion contains credible evidence from network operators, and that the extracted incident is supported by the discussion. They also verify whether the extracted fields are correct or can be reliably completed using external authoritative routing data.

Some incident reports explicitly provide all necessary fields. Others describe the anomaly in operational language but omit one or more fields. In such cases, experts consult auxiliary data sources, including historical BGP updates, IRR records~\cite{irr}, RPKI data~\cite{rpki}, and AS relationship datasets~\cite{caida-asrelation}. For example, when a report identifies a victim network and approximate event time but does not list all affected prefixes, experts use routing registry information to determine whether the missing field can be inferred. If a critical field remains ambiguous, the candidate is excluded from the final dataset. Moreover, we observe that operators often use terms such as ``route hijacking'' or ``BGP hijacking'' ambiguously, in some cases referring to prefix hijacking, and in others to path hijacking. Disambiguating these references requires contextual analysis. Therefore, experts are also responsible for verifying the anomaly type extracted from such operator descriptions.

For each retained incident, we annotate field-level provenance to distinguish reported fields from reconstructed ones. Disagreements between experts are resolved through discussion based on the original operator evidence. The entire validation process is completed within five business days ($\sim$32 hours). As a result, 232 high-confidence routing anomaly events are retained, representing 95.08\% of the initially extracted events. Among them, only 6 incidents require partial field reconstruction.
\section{Preliminary Study: Generalization Gap}\label{sec:preliminary}

Having constructed a verified benchmark, we next use it to revisit SOTA routing anomaly detection systems and general-purpose LLMs. This study serves two purposes. First, it examines whether existing detectors generalize to a broader set of real-world incidents (\cref{sec:soda_perf}). Second, it motivates the need to ground LLM reasoning in BGP protocol semantics and time-consistent routing evidence (\cref{sec:llm_perf}, \cref{sec:motivate_routellm}).

\subsection{Existing SOTA on the New Benchmark}\label{sec:soda_perf}

We first investigate whether existing SOTA routing anomaly detection systems generalize to the incidents in our benchmark.
We evaluate BEAM~\cite{beam2024}, a recent semantics-aware routing anomaly detector that outperforms prior learning-based approaches and reports perfect detection on 18 known incidents with only 5.53\% false alarms~\cite{beam2024}. 
BEAM learns AS ``routing roles'' from AS relationship data using a CNN-based representation learning framework and detects anomalies through abnormal AS-path changes.

We deploy the open-source implementation of BEAM~\cite{beamcodes}, retrain it using the latest CAIDA AS relationship dataset~\cite{caida-asrelation}, and follow the original evaluation setup.
Following BEAM, we collect BGP updates within twelve hours before and after each incident to preserve routing context and historical path information.
Table~\ref{tab:bg_models_acc} reports both event-level and message-level accuracy, where event-level accuracy measures whether an incident is detected and message-level accuracy measures the fraction of anomalous BGP updates correctly identified.

BEAM achieves only 57.67\% event-level accuracy and 36.71\% message-level accuracy on our benchmark, substantially lower than the results reported in the original paper.
We attribute this degradation to three factors.
First, AS relationship data alone may be insufficient to capture the full complexity of routing behaviors, and noisy relationship inferences can directly affect detection quality.
Second, the model may overfit to a limited set of historical anomaly patterns and generalize poorly to unseen incidents.
Third, BEAM's threshold-based path-difference scoring mechanism may be less robust under dynamically evolving routing conditions.

\subsection{General-purpose LLMs on the New Benchmark}\label{sec:llm_perf}

Recent studies have explored LLMs for cybersecurity and networking tasks such as traffic classification~\cite{trafficformer2025,guthula2023netfound}, protocol fuzzing~\cite{MRMM2024}, and penetration testing~\cite{pentestgpt2024}, demonstrating strong domain-adaptation capability through fine-tuning~\cite{XHWS2024}. 
Motivated by these advances, we investigate whether general-purpose LLMs can perform routing anomaly detection without routing-specific adaptation.
Using our benchmark, we evaluate five open-source LLMs, including DeepSeek-R1-Distill-Qwen-7B~\cite{qwen-7b-hgf}, Llama-3.1-8B-Instruct~\cite{llama3.1-8b-hgf}, QwQ-32B~\cite{qwq-32b-hgf}, DeepSeek-V3~\cite{ds-v3}, and DeepSeek-R1~\cite{ds-r1}, as well as three closed-source models: GPT-5, Gemini-3-Pro, and Claude-4.5-Haiku.
For each model, we prompt it to classify whether a BGP update is normal or anomalous and, if anomalous, identify its category.

Table~\ref{tab:bg_models_acc} summarizes the results.
DeepSeek-R1, Llama-3.1-8B-Instruct, and DeepSeek-R1-Distill-Qwen-7B achieve the best event-level accuracy among all tested LLMs, reaching 55.74\%, 54.24\%, and 53.45\%, respectively.
At the message level, Gemini-3-Pro achieves the highest accuracy of 42.61\%.
Notably, these results are obtained without any routing-domain adaptation and only use abnormal BGP updates, differing from the full evaluation setting in \cref{sec:eval}. Overall, some general-purpose LLMs can identify a subset of routing anomalies, but their performance varies across models and remains insufficient for reliable routing anomaly detection, motivating further routing-domain adaptation.

We further examine why some smaller LLMs achieve performance comparable to
or higher than larger models. We find no monotonic relationship between model
size and routing anomaly detection performance. For example, Llama-3.1-8B
achieves performance close to DeepSeek-R1 and consistently outperforms
QwQ-32B across anomaly types. Smaller instruction-tuned or distilled models
also exhibit higher output-format compliance and lower invalid-answer rates
than the larger reasoning-oriented models we evaluate. These results suggest
that model family, post-training, and response-format adherence, rather than
parameter scale alone, affect performance on this task.
  
\subsection{\sys Design Motivations}\label{sec:motivate_routellm}

The results above reveal a gap that is not addressed by either existing routing anomaly detectors or general-purpose LLMs alone.
Existing systems such as BEAM encode useful routing-domain knowledge, yet their performance drops substantially on our broader benchmark, suggesting limited generalization to diverse ASes, prefixes, time periods, and anomaly patterns.
In contrast, general-purpose LLMs can identify some routing anomalies, but their predictions remain inconsistent and unstable across models.

To better understand these limitations, we manually analyze representative false positives and false negatives.
We find that BEAM often fails when anomalies do not produce significant deviations from historical AS-path patterns or when incomplete AS relationship data makes suspicious paths appear normal.
LLMs fail for different reasons: they frequently treat AS paths and prefixes as ordinary text, confuse different anomaly types, and make decisions without grounding on time-dependent routing evidence such as prefix ownership, AS relationships, and historical origin behavior.
These observations indicate that general-purpose LLMs are insufficient as standalone classifiers for BGP updates.

Based on these findings, we derive three key design requirements for robust LLM-based routing anomaly detection:
\begin{itemize}
\item[R1] \textbf{BGP protocol semantics awareness.}
The detector should preserve the structured semantics of BGP data, rather than flattening prefixes, origin ASes, and AS paths into unstructured text.

\item[R2] \textbf{Routing-domain adaptability.}
The detector should incorporate routing-domain knowledge, including AS relationships and historical anomaly patterns, to reason about path plausibility, prefix ownership, and route-leak policies.

\item[R3] \textbf{Time-consistent routing evidence grounding.}
The detector should ground its decisions on routing evidence consistent with the event timestamp, since routing knowledge continuously evolves over time.
\end{itemize}

\section{\sys Design}\label{sec:design}

To address the three requirements identified in~\cref{sec:motivate_routellm}, we design \sys, a BGP-aware and evidence-grounded LLM framework for routing anomaly detection.
\sys consists of three core components: BGP-semantic-aware tokenization, routing-domain adaptation, and time-aware routing knowledge retrieval, corresponding to \textbf{R1}, \textbf{R2}, and \textbf{R3}, respectively.
Instead of treating BGP updates as plain text, \sys grounds LLM reasoning on protocol-structured routing fields and retrieved routing evidence, including historical BGP announcements, AS relationships, and IRR/RPKI records.
This design enables \sys to generalize beyond the limited incident sets used by prior detectors while reducing the instability observed in general-purpose LLMs.
The following sections describe the design of each component in detail: BGP-semantic-aware tokenization (\cref{sec:token_design}), routing-domain adaptation (\cref{sec:fine_tune_design}), and time-aware routing knowledge retrieval (\cref{sec:rag_design}).

\subsection{BGP-semantic-aware Tokenization}\label{sec:token_design}
\sys uses a BGP-semantic-aware tokenizer to encode routing anomaly detection inputs for LLMs. Trained on BGP data and textual detection instructions, the tokenizer captures BGP semantics while retaining natural language understanding, enabling LLMs to effectively process both structured routing data and unstructured text.

To achieve this, the tokenizer must treat critical BGP elements, such as AS paths and attributes, as atomic semantic units to maintain their intrinsic meaning and support \textit{protocol-level} segmentation, \eg splitting ``AS\_PATH: 2914 7018'' into structured tokens.
These help achieve the routing domain adaptation to address limitations of the generic tokenizer in handling numeric sequences, \eg over-splitting AS numbers like 3356 into 3, 3, 5, 6, and protocol keywords, \eg incorrectly tokenizing ``next-hop'' as ``next'' and ``hop''.

To build this tokenizer, we prepare a dataset of BGP announcements (both abnormal and normal, as described in~\cref{sec:eval}) paired with 200 manually written detection instructions of varied phrasing. Each training instance concatenates an instruction, a BGP update prefixed by a special \texttt{<BGP Update:>} token, and the expected output, yielding 22,800 samples.
We show the instruction example of the training data in Fig.~\ref{fig:data4token}. We use the whole information within the BGP announcements as the input, instead of specifying predefined features, making the model learn the features automatically.
\begin{figure}[t]
\centering
\begin{lstlisting}
(*\bfseries Instruction*): Please analyze the BGP routing update information following <BGP Update:> to determine whether the update is normal. If any anomalies are detected, please clearly identify the type of anomaly including prefix hijack, path hijack, or route leak. If you do not determine any anomaly type, anomaly_type is filled in as none.
(*\bfseries <BGP Update:>*) TIME: 09/03/24 01:47:43, TYPE: BGP4MP/MESSAGE/Update, FROM: 2406:840:eb83::1 AS140731, ORIGIN: IGP, ASPATH: 140731 137990 ...
(*\bfseries Output*): none.
\end{lstlisting}
\vspace{-3mm}
\caption{An example of the training data for the BGP-semantic-aware tokenizer.}
\vspace{-5mm}
\label{fig:data4token}
\end{figure}

\parab{Tokenizer training.}
Using the prepared training data, we train the SentencePiece model~\cite{sentencepiece2018} to generate the specialized tokenizer.
We choose it because the SentencePiece model can integrate seamlessly into any LLM's vocabulary without architectural changes and reliably preserves rare but critical routing tokens on our moderate-sized BGP corpus.
We train the model by byte pair encoding (BPE) method.
Different from unigram language model (ULM), the training using BPE is more greedy and does not explicitly minimize a probabilistic loss~\cite{bpevsulm}.
Besides, because LLMs rarely have a large amount of BGP data to train, we can see BGP data as a new ``natural language'', and directly use the tokens from the specialized tokenizer to extend the vocabulary of LLMs, and assign initial embedding values to the extended tokens and update them in the fine-tuning stage.

\subsection{Routing-domain Adaptation}\label{sec:fine_tune_design}

Although LLMs are aware of routing anomaly concepts (e.g., prefix hijacking, route leaks, valley-free policies), they cannot apply this knowledge in practice. We attribute this gap to two factors: (i) dynamic, domain-specific information such as AS relationships and prefix ownership is absent from standard pretraining corpora, and (ii) LLMs have never been exposed to real BGP announcements and thus lack the practical experience needed for detection. To bridge this gap, we adopt a two-stage fine-tuning strategy.

In the first stage, we fine-tune the LLM on AS relationship data using a masked-token reconstruction objective. By randomly masking tokens in statements such as “AS1 is a [MASK] AS of AS2” and training the model to reconstruct the masked routing-relation tokens, the model learns AS-level relationship semantics. In the second stage, we build on the resulting model and fine-tune it with normal and abnormal BGP announcements 
using a standard cross-entropy classification loss. Crucially, we apply LoRA~\cite{lora2021} in this stage: since the goal is to teach the model how to \emph{apply} already acquired routing knowledge rather than learn entirely new concepts, updating only low-rank adapters is sufficient and reduces training overhead. 

\subsection{Time-aware Routing Knowledge Retrieval}\label{sec:rag_design}

Routing anomaly decisions often depend on time-dependent evidence. A prefix-origin pair may appear normal only if the origin had historically announced the prefix; an AS path may indicate a route leak only when viewed under the AS relationships valid at that time. LLMs lack access to such dynamic knowledge, and post-event data can mislead detection.

To address this, \sys integrates a time-aware routing knowledge retriever based on Retrieval-Augmented Generation (RAG)~\cite{ROLY2023}. The retriever organizes timestamped routing evidence into an external knowledge base, including prefix-origin histories, AS relationship snapshots, and RPKI/IRR records. It indexes all entries for semantic retrieval. Given an input BGP update, \sys extracts its prefix, origin AS, AS path, and timestamp, and retrieves the top-K most relevant evidence records that are temporally consistent with the event. For example, for a prefix announcement, it may return the set of origins that announced the prefix before the event, the RPKI status at event time, and previously observed AS-level adjacencies. The retriever formats the selected evidence as an evidence block and appends it to the model input. We show an example of a time-aware evidence block in Fig.~\ref{fig:evidence_example}.

\begin{figure}[t]
\centering
\begin{lstlisting}
<EVIDENCE>
(*\bfseries Historical origins of 50.92.0.0/17 before YYYYMMDD*):
  ASXXXX observed in RouteViews and RIS.
(*\bfseries Observed origin in current update*):
  AS10990.
(*\bfseries AS path evidence*):
  AS1299-AS7219 observed before;
  AS7219-AS10990 rarely observed before event time.
(*\bfseries RPKI status*):
  no matching valid ROA found at event time.
</EVIDENCE>
\end{lstlisting}
\vspace{-1mm}
\caption{An example of a time-aware evidence block appended to the model input.}
\label{fig:evidence_example}
\vspace{-5mm}
\end{figure}

This retriever does not model the temporal dynamics of BGP update sequences or replace time-series modeling. Instead, it provides external routing evidence that is consistent with the timestamp of the analyzed update. By this method, \sys reduces its dependence on static model parameters and avoids using outdated routing information. This design also allows newly collected routing data to be incorporated without full retraining.

After augmenting the input with time-consistent evidence, \sys generates an anomaly decision in a structured and evidence-grounded format. The model outputs both an anomaly label and the routing evidence used to support the prediction, which encourages reasoning grounded in retrieved facts rather than memorized patterns. 

\section{System Implementation}\label{sec:sys_imple}

\begin{figure}[t]
  \centering
  \includegraphics[width=1.00\columnwidth]{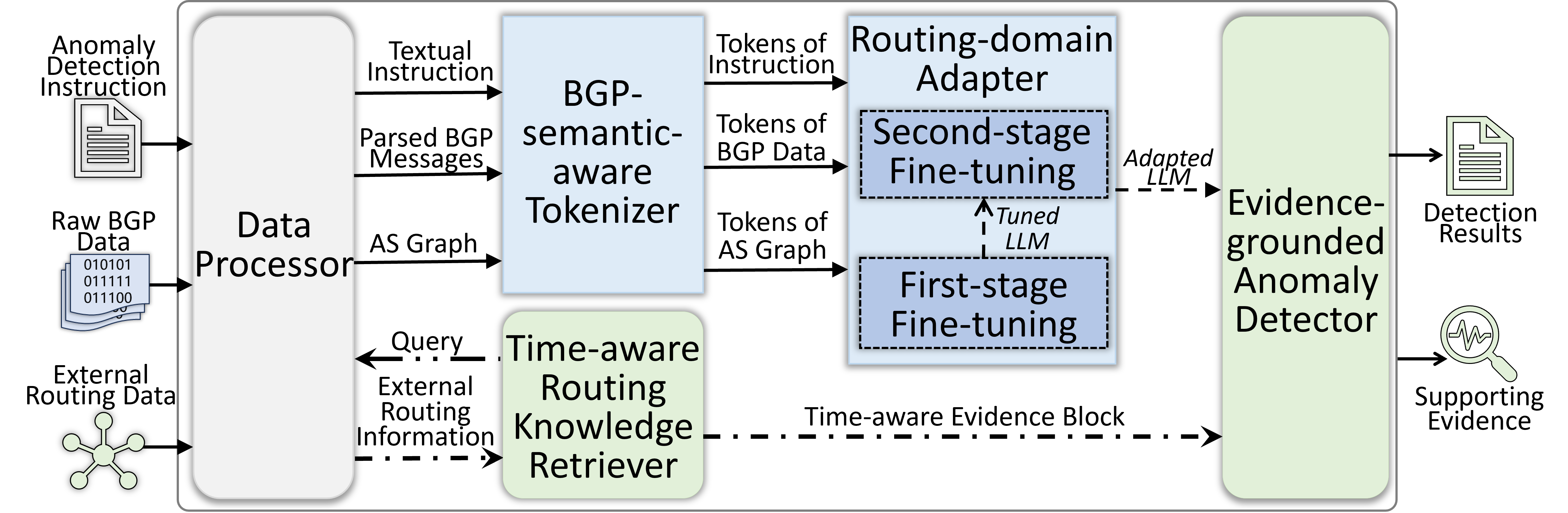}
  \vspace{-5mm}
  \caption{Overview of \sys system. The BGP-semantic-aware Tokenizer and Routing-domain Adapter are used during the training and fine-tuning stage to construct a routing-adapted LLM, while Time-aware Routing Knowledge Retriever and Evidence-grounded Anomaly Detector are used during inference for routing anomaly detection.}
  \vspace{-5mm}
  \label{fig:routellm_overview}
\end{figure}

Fig.~\ref{fig:routellm_overview} illustrates the overall architecture of \sys, which consists of
five modules. The \textit{data processor} converts raw BGP data into
human-readable representations, translates AS relationships into natural
language, and prepares detection instructions. The
\textit{BGP-semantic-aware tokenizer} extends the native LLM tokenizer
with routing-aware tokens while retaining the original tokenizer for
natural-language instructions. The \textit{routing-domain adapter}
performs AS-relationship fine-tuning followed by LoRA-based fine-tuning
on labeled BGP updates. During inference, the
\textit{time-aware routing knowledge retriever} retrieves
timestamp-consistent prefix-origin, AS-relationship, RPKI/IRR, and
historical BGP evidence, which the \textit{evidence-grounded anomaly
detector} combines with the BGP update and detection instruction to
predict the anomaly type and supporting evidence.

We implement \sys using Python 3.12 and PyTorch 2.1.2. Data collection
from RouteViews and RIS requires approximately 307 LOC, while BGP parsing
and fine-tuning-data preparation require approximately 22 and 201 LOC,
respectively. The tokenizer, routing-domain adapter, retriever, and anomaly
detector comprise approximately 1,553 LOC. We use this pipeline to
fine-tune Llama-3.1-8B-Instruct and use the resulting model
for the experiments in~\cref{sec:eval}. We do not use closed-source API-based
LLMs as the \sys foundation because they do not support the required
low-level model customization.
\section{Evaluation}\label{sec:eval}
We evaluate \sys from four aspects: overall routing
anomaly detection performance, sensitivity to the normal-to-abnormal message ratio, contributions of key model-adaptation
components, and training and inference efficiency.

\begin{table*}[t]
\centering
\caption{Overall routing anomaly detection performance across five event-level train/test splits. Results are reported as means, with standard
deviations in parentheses.}
\label{tab:overall_performance}

\setlength{\tabcolsep}{3.5pt}
\renewcommand{\arraystretch}{1.05}

\begin{minipage}[t]{0.49\textwidth}
\centering
\textbf{(a) Event-level Performance}

\begin{tabular}{c||c|c|c|c|c}
\hline\hline
{Metric} & {\sys} & {BEAM} &
{MS-LSTM} & {RF} & {GCN} \\ \hline

AC
& \makecell{\textbf{0.8713}\\(0.0118)}
& \makecell{0.5611\\(0.0197)}
& \makecell{0.3797\\(0.0213)}
& \makecell{0.4873\\(0.0182)}
& \makecell{0.5124\\(0.0165)} \\ \hline

PR
& \makecell{\textbf{0.8420}\\(0.0143)}
& \makecell{0.5532\\(0.0211)}
& \makecell{0.3310\\(0.0232)}
& \makecell{0.4680\\(0.0198)}
& \makecell{0.4887\\(0.0181)} \\ \hline

RC
& \makecell{\textbf{0.8795}\\(0.0130)}
& \makecell{0.5924\\(0.0201)}
& \makecell{0.3568\\(0.0221)}
& \makecell{0.5012\\(0.0189)}
& \makecell{0.5256\\(0.0173)} \\ \hline

F1
& \makecell{\textbf{0.8571}\\(0.0138)}
& \makecell{0.5619\\(0.0206)}
& \makecell{0.3297\\(0.0221)}
& \makecell{0.4751\\(0.0190)}
& \makecell{0.5035\\(0.0170)} \\ \hline

FPR
& \makecell{\textbf{0.0735}\\(0.0085)}
& \makecell{0.2134\\(0.0150)}
& \makecell{0.3287\\(0.0181)}
& \makecell{0.2585\\(0.0142)}
& \makecell{0.1913\\(0.0131)} \\ \hline

FNR
& \makecell{\textbf{0.1205}\\(0.0130)}
& \makecell{0.4076\\(0.0201)}
& \makecell{0.6432\\(0.0221)}
& \makecell{0.4988\\(0.0189)}
& \makecell{0.4744\\(0.0173)} \\ \hline\hline

\end{tabular}
\end{minipage}
\hfill
\begin{minipage}[t]{0.49\textwidth}
\centering
\textbf{(b) Message-level Performance}

\begin{tabular}{c||c|c|c|c|c}
\hline\hline
{Metric} & {\sys} & {BEAM} &
{MS-LSTM} & {RF} & {GCN} \\ \hline

AC
& \makecell{\textbf{0.9465}\\(0.0089)}
& \makecell{0.4263\\(0.0131)}
& \makecell{0.5169\\(0.0179)}
& \makecell{0.4862\\(0.0148)}
& \makecell{0.5618\\(0.0141)} \\ \hline

PR 
& \makecell{\textbf{0.8976}\\(0.0115)}
& \makecell{0.2531\\(0.0180)}
& \makecell{0.4396\\(0.0203)}
& \makecell{0.4187\\(0.0171)}
& \makecell{0.4876\\(0.0157)} \\ \hline

RC 
& \makecell{\textbf{0.9212}\\(0.0096)}
& \makecell{0.2018\\(0.0178)}
& \makecell{0.3591\\(0.0189)}
& \makecell{0.4015\\(0.0158)}
& \makecell{0.4698\\(0.0155)} \\ \hline

F1 
& \makecell{\textbf{0.8997}\\(0.0102)}
& \makecell{0.2078\\(0.0189)}
& \makecell{0.3691\\(0.0199)}
& \makecell{0.3996\\(0.0161)}
& \makecell{0.4701\\(0.0157)} \\ \hline

FPR 
& \makecell{\textbf{0.0215}\\(0.0043)}
& \makecell{0.2487\\(0.0129)}
& \makecell{0.1914\\(0.0148)}
& \makecell{0.1769\\(0.0119)}
& \makecell{0.1482\\(0.0103)} \\ \hline

FNR 
& \makecell{\textbf{0.0788}\\(0.0096)}
& \makecell{0.7982\\(0.0178)}
& \makecell{0.6409\\(0.0189)}
& \makecell{0.5985\\(0.0158)}
& \makecell{0.5302\\(0.0155)} \\ \hline\hline

\end{tabular}
\end{minipage}

\end{table*}

\parab{Summary of Results}:
\begin{icompact}
\item Across five event-level splits, \sys achieves 87.13\% event-level and 94.65\% message-level accuracy, outperforming the strongest baselines by 55.30\% and 68.50\%, respectively.
\item \sys remains robust across normal-to-abnormal message ratios from 100:1 to 10000:1: recall decreases only from 0.9276 to 0.9119, while the FPR remains close to 0.02.
\item We show that the BGP-semantic-aware tokenizer improves all message-level metrics by at least 10.15\%, while also enhancing BGP data processing efficiency.
\item We validate the effectiveness of the two-stage fine-tuning in \sys. Fine-tuning with AS relationship data alone improves
message-level metrics by at least 51.26\%, while fine-tuning with BGP data alone improves them by at least 86.50\%.
\item \sys incurs moderate offline training cost while achieving practical inference efficiency.
\end{icompact}

\parab{Testbed.}
We use two servers, each with 4$\times$ NVIDIA A6000 GPUs, an Intel Xeon Icelake CPU, and 400GB RAM, running Ubuntu 22.04 LTS, PyTorch 2.1.2, and CUDA 12.1.

\noindent\textbf{Dataset.}
We curate a dataset of 249 operator-confirmed routing anomaly events.
Among them, 232 events are newly collected from public mailing lists (\cref{sec:dataset}), and 17 are from the 18 incidents used in BEAM~\cite{beam2024} (with one overlap).
For these events, we collect 68{,}913 abnormal BGP update messages, each linked to a documented incident.
Because real-world BGP routing data is highly imbalanced and anomalous updates constitute only a tiny fraction of operational traces~\cite{DYLQ2021}, we sample normal messages with an average ratio of 1000:1.
Normal messages are sampled from the same temporal window and related routing context as the corresponding abnormal messages. The 1000:1 ratio therefore refers to the message ratio within each anomaly event.

\noindent\textbf{Train/test split.}
We conduct five randomized event-level train/test splits, each containing 170 training events and 79 test events. We stratify the splits by anomaly category to preserve similar category distributions. All BGP messages associated with the same event remain in the same split to avoid data leakage. We use a subset of the training events for validation.

\parab{Baselines}.
We compare \sys with four categories of methods:
(1) \textbf{BEAM}~\cite{beam2024}, a routing anomaly detector that incorporates AS relationship knowledge through AS embeddings and path-difference scoring. For each randomized split, we retrain its AS-representation components using the routing data corresponding to the evaluation period. Components fixed by the original BEAM methodology remain unchanged. 
(2) \textbf{MS-LSTM}~\cite{mslstm2016}, a time-series learning approach that models historical BGP behavior through multi-scale LSTMs. We implement it following the original paper and train it using the same event-level training split as \sys. 
(3) \textbf{Random Forest} (RF), a representative machine-learning approach previously evaluated for BGP anomaly detection using routing features~\cite{SOFS2019}. We train RF using routing-context features, including AS-path length, origin changes, prefix length, newly observed AS-path edges, and historical prefix-origin behavior.
(4) \textbf{Graph Convolutional Network} (GCN), a representative graph-based approach previously applied to BGP anomaly detection by modeling AS-level routing topology~\cite{wu2025graphbgp}. We construct the AS-level topology from BGP observations and use GCN to learn graph representations for classifying BGP updates.
All trainable baselines use the same randomized event-level train/test splits as \sys and are evaluated on the same test events.

\parab{Metrics}.
We evaluate routing anomaly detection at both event and message levels.
We report Accuracy (AC), Precision (PR), Recall (RC), F1 score,
false-positive rate (FPR), and false-negative rate (FNR) at both levels.

\subsection{Overall Performance}\label{sec:eval_overall_accu}
Table~\ref{tab:overall_performance} summarizes the performance of \sys and
the compared schemes across five randomized event-level train/test splits.
We report the mean and standard deviation for both event-level and
message-level metrics.
At the event level, \sys achieves an average accuracy of 0.8713, compared
with 0.5611 for BEAM, the strongest baseline, corresponding to a 55.3\%
relative improvement. It also achieves the highest precision, recall, and
F1 score of 0.8420, 0.8795, and 0.8571, while reducing the FPR and FNR to
0.0735 and 0.1205, respectively.
At the message level, \sys achieves an accuracy of 0.9465 and an F1 score
of 0.8997, compared with 0.5618 and 0.4701 for GCN, the strongest baseline
at this level. It further reduces the FPR and FNR to 0.0215 and 0.0788.
The small standard deviations across the five splits indicate that these
performance gains remain stable under different train/test partitions.

\vspace{-2mm}

\subsection{Impact of Normal-to-Abnormal Sampling Ratio}
\label{sec:ratio_sensitivity}

\begin{table}[t]
\centering
\caption{Message-level anomaly detection performance under different normal-to-abnormal message sampling ratios.}
\label{tab:ratio_sensitivity}
\scalebox{0.95}{
\begin{tabular}{c||c|c|c|c|c|c}
\hline\hline
Ratio   & AC     & PR     & RC     & F1     & FPR    & FNR \\ \hline
100:1   & 0.9401 & 0.9139 & 0.9276 & 0.9112 & 0.0207 & 0.0724 \\ \hline
1000:1  & 0.9465 & 0.8976 & 0.9212 & 0.8997 & 0.0215 & 0.0788 \\ \hline
10000:1 & 0.9492 & 0.8743 & 0.9119 & 0.8846 & 0.0228 & 0.0881 \\ 
\hline\hline
\end{tabular}
}
\end{table}

We further evaluate the sensitivity of \sys to the normal-to-abnormal
message ratio. For each split, we keep the trained model and anomalous messages
unchanged and vary only the number of sampled normal messages, using ratios
of 100:1, 1000:1, and 10000:1. Table~\ref{tab:ratio_sensitivity} summarizes
the results.

\sys remains robust across different sampling ratios. As the
ratio increases from 100:1 to 10000:1, recall decreases only from 0.9276 to
0.9119, while the FPR remains close to 0.02. Precision and F1 decrease
moderately from 0.9139 and 0.9112 to 0.8743 and 0.8846, respectively, under
the more extreme imbalance. Meanwhile, accuracy remains stable. These
results show that \sys maintains reliable anomaly detection performance
under substantial variations in the normal-to-abnormal message ratio.

\subsection{Ablation Study}\label{sec:ablation_study}
We next investigate the
contribution of the two key model-adaptation components in \sys: the
BGP-semantic-aware tokenizer and the two-stage fine-tuning strategy.
For the ablation experiments, we use a fixed train/test split and
keep all other experimental settings unchanged. 

\parab{BGP-semantic-aware tokenizer.}
We evaluate the tokenizer from two perspectives: detection performance and token efficiency. 
For token-efficiency evaluation, we randomly sample 200 instruction--BGP data pairs and compare the resulting token lengths between \sys and five open-source LLMs.
\begin{figure}[t]
  \centering
    \includegraphics[width=0.85\columnwidth]{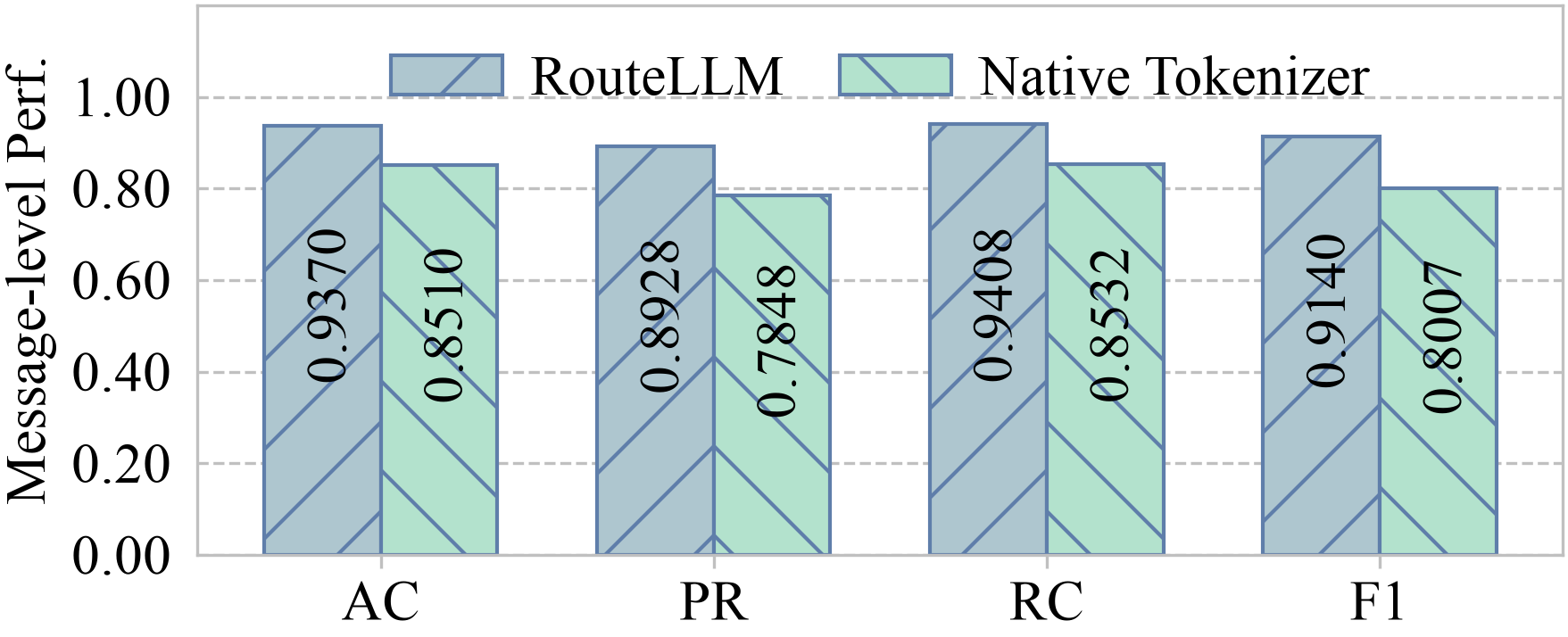}
    \vspace{-2mm}
    \caption{Message-level performance of \sys with and without the BGP-semantic-aware tokenizer.}
    \vspace{-5mm}
  \label{fig:eval_token_perf}
\end{figure}
Fig.~\ref{fig:eval_token_perf} compares \sys using the specialized tokenizer against the native tokenizer of the underlying LLM, while keeping all other settings unchanged.
The specialized tokenizer improves all message-level metrics by at least 10.15\%, demonstrating its effectiveness in capturing routing semantics.

\begin{figure}[t]
  \centering
    \includegraphics[width=0.85\columnwidth]{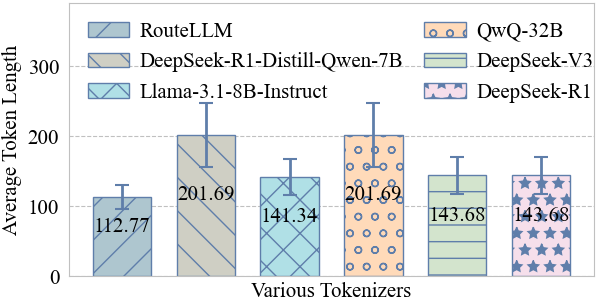}
    \vspace{-2mm}
    \caption{Average token lengths of \sys and baseline LLMs on the same BGP inputs.}
    \vspace{-5mm}
  \label{fig:eval_token_len}
\end{figure}

Fig.~\ref{fig:eval_token_len} further shows that \sys produces substantially shorter token sequences, with an average token length of 112.77, compared to 141.34--201.69 for baseline LLMs.
This reduction comes from routing-aware tokenization tailored to BGP structures, improving both representation compactness and inference efficiency.
As shown in Fig.~\ref{fig:eval_latency}, this tokenizer contributes to a 76.42\% latency reduction compared with Llama-3.1-8B-Instruct.

\parab{Two-stage fine-tuning.}
We next evaluate the contribution of each fine-tuning stage by separately enabling AS-relationship fine-tuning and BGP-data fine-tuning.
\begin{figure}[t]
  \centering
    \includegraphics[width=0.85\columnwidth]{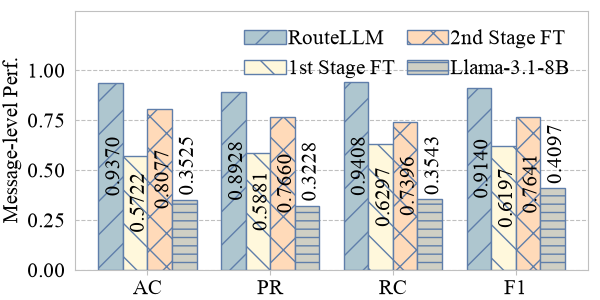}
    \vspace{-2mm}
    \caption{Message-level performance of \sys with different fine-tuning strategies.}
  \label{fig:as_fine_tuning}
\end{figure}
Fig.~\ref{fig:as_fine_tuning} compares full \sys, \sys with only AS-relationship fine-tuning, \sys with only BGP-data fine-tuning, and the original Llama-3.1-8B-Instruct.

Using only AS-relationship fine-tuning significantly improves message-level performance over the original LLM, achieving gains of 62.33\% in accuracy, 82.19\% in precision, 77.73\% in recall, and 51.26\% in F1 score.
Besides, fine-tuning only on labeled BGP data improves message-level accuracy, precision, recall, and F1 score by 129.13\%, 137.30\%, 108.75\%, and 86.50\%, respectively.
The performance gap between this setting and full \sys further demonstrates that AS-relationship knowledge and anomaly-specific training are complementary, and combining them yields the best overall performance.

\subsection{Training and Inference Efficiency}\label{sec:eval_overhead}
\begin{table}[t]
\centering
\caption{Training time of the BGP-semantic-aware tokenizer and the two fine-tuning stages.}
\vspace{-2mm}
\label{tab:eval_time_overhead}
\scalebox{0.82}{
\begin{tabular}{c||c|c|c}
\hline\hline
\begin{tabular}[c]{@{}c@{}}Time Overhead\\ (Hours)\end{tabular}        & \begin{tabular}[c]{@{}c@{}}Tokenizer\\ Training\end{tabular} & \begin{tabular}[c]{@{}c@{}}First Fine-tuning Stage\\ (Full Fine-tuning)\end{tabular} & \begin{tabular}[c]{@{}c@{}}Second Fine-tuning Stage\\ (LoRA-based Fine-tuning)\end{tabular} \\ \hline
\begin{tabular}[c]{@{}c@{}}Average Time \\ for Each Epoch\end{tabular} & -                  & 307.45                                                                               & 12.06                                                                                       \\ \hline
Total Time                                                             & 3.97               & 309.10                                                                               & 36.98                                                                                       \\ \hline\hline
\end{tabular}
}
\end{table}

\begin{figure}[t]
  \centering
    \includegraphics[width=0.85\columnwidth]{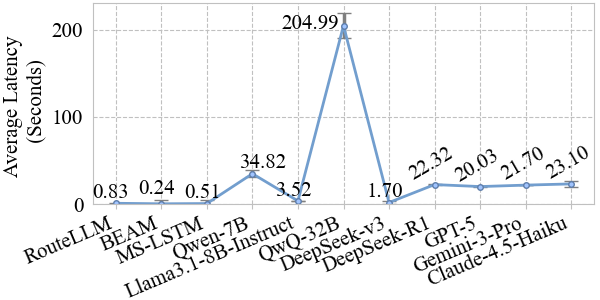}
    \caption{Average detection latency of \sys and the compared schemes. ``Qwen-7B'' denotes ``DeepSeek-R1-Distill-Qwen-7B''.}
  \label{fig:eval_latency}
  \vspace{-7mm}
\end{figure}

Table~\ref{tab:eval_time_overhead} reports the offline training cost of \sys. Tokenizer training
requires 3.97 hours, while the first-stage full fine-tuning and
second-stage LoRA-based fine-tuning require 309.10 and 36.98 hours,
respectively. The higher cost of the first stage results from updating
the full model, whereas LoRA substantially reduces the cost of
anomaly-specific adaptation.

Fig.~\ref{fig:eval_latency} compares inference latency. BEAM has the lowest latency due to
its lightweight feature-matching design, while reasoning-oriented LLMs
incur substantially higher latency due to longer generations. \sys
achieves sub-second to near-second latency per BGP update and is faster
than the evaluated closed-source reasoning models. Since routing anomaly
detection operates on BGP update streams and operational events rather
than per-packet forwarding, this latency is sufficient for operational
monitoring and incident analysis.
\section{Related Work}\label{sec:related}

Routing anomaly detection can be broadly grouped into three categories.

\vspace{-1mm}
\parab{Statistical behavior-based routing anomaly detection.}
Statistical approaches detect routing anomalies using manually designed
heuristics over BGP features~\cite{artemis2018,LJDD2005,bgpmon2009,ZCJL2007,HXMZ2007,buddyguard2012,VPAT2015,TCRP2019,HTAT2024,ispy2008,heap2016,argus2012}, such as AS-path length, prefix
reachability, and historical AS behavior. They operate on control-plane,
data-plane, or hybrid routing signals~\cite{VPAT2015,heap2016,HXMZ2007,argus2012}, but often require manual tuning, extensive historical
data, or costly infrastructure, and remain constrained by scarce
ground-truth datasets.

\vspace{-1mm}
\parab{Machine learning-based routing anomaly detection.}
Learning-based approaches infer abnormal routing patterns directly from
BGP data~\cite{beam2024,mslstm2016,DYLQ2021,rqa2015,ARNM2012,LABM2014,DSTM2009,TGTO2012,STSY2020,HKTP2021,ap2vec2022,SOFS2019,wu2025graphbgp}. For example, BEAM~\cite{beam2024} learns AS embeddings
from AS-relationship data and detects abnormal path changes, while
MS-LSTM~\cite{mslstm2016} models temporal dependencies in BGP update streams.
These methods depend heavily on labeled anomaly data and have limited
ability to incorporate broader routing-domain knowledge.

\vspace{-1mm}
\parab{Applying LLMs in Security and Networking.}
LLMs have recently been applied to security and networking tasks, including
intrusion detection~\cite{HPSP2024,guthula2023netfound}, protocol fuzzing~\cite{MRMM2024}, penetration
testing~\cite{pentestgpt2024}, network management~\cite{mani2023enhancing,wang2024netconfeval,wang2025intent}, network
testing~\cite{singha2025extremal}, and network-function optimization~\cite{ghasemirahni2024deploying}.
Several studies also evaluate LLM networking capabilities using
domain-specific benchmarks~\cite{anwar2025understanding}. 
However, their application to routing anomaly detection remains largely
unexplored. We investigate this gap through BGP-semantic-aware
tokenization, routing-domain adaptation, and evidence grounding.
\section{Limitations and Future Work}\label{sec:limit}

Although \sys is effective for boosting routing anomaly detection, it still has several limitations that require further investigation. We discuss them below.

\noindent\textbf{Dataset representativeness.}
Our benchmark is not a statistically representative sample of all routing anomalies on the Internet. It covers incidents that were observed in the operator mailing lists. Anomalies that were not observed or publicly reported may therefore be absent, which may limit the generalizability of the evaluation results.

\noindent\textbf{Reporting and regional biases.}
Incidents with greater operational impact or visibility are more likely to be reported, while regions and operator communities with more active mailing lists may be better represented. Incidents discussed through private channels not covered by our data may also be missing.

\noindent\textbf{Agent-based evidence collection.}
Some routing anomaly decisions may require information beyond the evidence currently indexed by \sys.
An agent-based extension could query additional sources, such as WHOIS, looking glasses, and operator reports. 
Such an approach may provide richer evidence for complex incidents, but also introduces challenges in tool selection, external-service availability, and inference latency. We leave this exploration for future work.

\noindent\textbf{Model generality.}
We implement and evaluate \sys using Llama-3.1-8B-Instruct as the base model. Although the design is model-agnostic, the current evaluation does not establish that other model families would achieve the same improvements after routing-domain adaptation. 
Evaluating \sys with additional foundation models is an important direction for future work.

\section{Conclusion}\label{sec:conclu}
We collect an unprecedented amount of routing incident data from public
operator mailing lists with the help of LLMs and construct a verified
benchmark containing 232 real-world routing anomaly events. We further
develop \sys, an LLM-based routing anomaly detection system that integrates
BGP-semantic-aware tokenization, routing-domain adaptation, and time-aware
routing evidence retrieval. Experimental results show that \sys achieves
87.13\% event-level accuracy and 94.65\% message-level accuracy,
substantially outperforming the compared baselines.

\section*{Acknowledgments}
We thank the IEEE ICNP 2026 shepherd and the anonymous reviewers for their constructive comments. Li Chen is the corresponding author.
This work was supported by National Cyber Security-National Science and Technology Major Project (No. 2025ZD1502400).

\clearpage

\balance
\bibliographystyle{IEEEtran}
\bibliography{reference}
\clearpage
\newpage
\appendix
\subsection{Ethical Considerations}\label{sec:appendix_ethic}
Our work addresses ethical aspects related to dataset construction, LLM fine-tuning, and experimental evaluation, as detailed below.

\parab{Dataset construction.} 
The fine-tuning dataset is constructed using AS relationship data, BGP data, and natural language instructions. Both the AS and BGP data originate from open-source datasets. The natural language instructions are generated through a combination of manual annotation and AI assistance, with strict adherence to avoiding any corporate confidential or private information. All routing anomaly events are collected from publicly accessible sources. Any organizational names, personal names, email addresses, IP addresses, or AS numbers referenced in Figs~\ref{fig:exp_abnormal_report},~\ref{fig:evidence_example},~\ref{fig:append_detailed_exp_abnormal_report_1}, and \ref{fig:append_detailed_exp_abnormal_report_2} are already public within these mailing lists. 
This dataset has been reviewed and approved by our institutional ethics review board (IRB) to ensure ethical integrity and compliance.

\parab{LLM fine-tuning.}
The fine-tuning of the LLM is performed exclusively on our local servers to ensure security and avoid any adverse impact on external systems or parties.

\parab{Experimental evaluation.}
Evaluation experiments for \sys, BEAM, MS-LSTM, DeepSeek-R1-Distill-Qwen-7B, and Llama-3.1-8B-Instruct are conducted locally on our servers. Experiments involving QwQ-32B, DeepSeek-V3, DeepSeek-R1, GPT-5, Gemini-3-Pro, and Claude-4.5-Haiku are carried out via official APIs provided by the respective companies, in full compliance with their terms of service. All experimental procedures have been disclosed to and approved by our IRB.

\subsection{Extended Background and Motivation}\label{sec:appendix_ext_bg}
\subsubsection{Routing Anomaly Categories}\label{sec:appendix_anomaly_cat}
In this work, we focus on the three major categories of routing anomalies: {\em prefix hijacking}, {\em path hijacking}, and {\em route leak}.
Prefix hijacking occurs when an AS maliciously announces IP prefixes that it does not own.
Path hijacking refers to a scenario in which an AS legitimately announces its own prefixes but maliciously manipulates the AS path to intercept traffic intended for specific networks.
Route leak happens when an AS unintentionally propagates BGP routes to unauthorized ASes, thereby violating established routing policies.

Please note that this work does not attempt to distinguish legitimate Multi-Origin AS (MOAS) announcements from routing anomalies, as ground-truth labels for legitimate MOAS cannot be obtained from public mailing lists. We leave the classification of legitimate MOAS to future work.

\subsubsection{Limitations of Routing Anomaly Data from Public Anomaly Detection Systems}\label{sec:limit_existing_dec}
We conduct an auxiliary experiment in which we fine-tune \sys using
incident data obtained from Cloudflare Radar~\cite{cloudflare}, while
keeping the model architecture and training procedure unchanged.
For the dataset from Cloudflare Radar used for fine-tuning the model,
we collect all the incidents via Cloudflare Radar's API and collect all the corresponding BGP data relating to these incidents to construct the training data.
Table~\ref{tab:cloudflare} shows the results.
We can see, when evaluated against our manually verified dataset of real anomalies (\cref{sec:dataset}), the model performed poorly.
This significant performance gap demonstrates that models trained on such data fail to generalize to real-world scenarios and are ineffective for accurate anomaly detection.

\begin{table}[t]
\centering
\caption{The routing anomaly detection performance of \sys trained using manually verified dataset of real anomalies (\cref{sec:dataset}) and dataset from Cloudflare Radar~\cite{cloudflare_radar}, respectively.}
\label{tab:cloudflare}
\scalebox{0.95}{
\begin{tabular}{c||c|c}
\hline\hline
Metrics & \begin{tabular}[c]{@{}c@{}}\sys\\ (\textbf{Verified Dataset})\end{tabular} & \begin{tabular}[c]{@{}c@{}}\sys\\ (\textbf{Dataset from Cloudflare Radar})\end{tabular} \\ \hline
AC      &           0.9370           &                               0.6715                                                                                   \\ \hline
PR      &           0.8928                                                                                   &                        0.5832                                                                                    \\ \hline
RC      &           0.9408                                                                                   &                        0.6299                                                                                    \\ \hline
F1      &           0.9140                                                                                     &                      0.4326                                                                                   \\ \hline\hline
\end{tabular}
}
\end{table}

\subsubsection{Examples of Routing Anomaly Incident Reports}\label{sec:appendix_exp_nanog}

Fig.~\ref{fig:append_detailed_exp_abnormal_report_1} shows the examples for BGP prefix hijacking and route leak event reports from NANOG mailing list~\cite{hijackexp,routeexp}, and Fig.~\ref{fig:append_detailed_exp_abnormal_report_2} shows the example for BGP path hijacking event report.

\subsection{Extended Routing Anomaly Data Collection}\label{sec:appendix_data_collect}
\subsubsection{Challenges for Extracting Incidents from Public Mailing Lists}\label{sec:challenges4parse}
To illustrate the challenges associated with collecting routing anomaly incidents from public mailing lists, we examine real-world examples drawn from NANOG's routing incident reports. 
Our analysis uses two BGP hijacking incidents and one route leak event,
which we present in Figs.~\ref{fig:append_detailed_exp_abnormal_report_1} and~\ref{fig:append_detailed_exp_abnormal_report_2}, as examples to introduce the challenges.

We summarize the challenges below.
\begin{itemize}
\item[C1] \textbf{Unstructured Data}: The emails in the mailing lists typically consist of unstructured operational data, including raw traceroute outputs as shown in Fig.~\ref{fig:app_exp_leak} and device command results, \eg ``\texttt{show route 128.10.4.0/24 detail}'' shown in Fig.~\ref{fig:append_detailed_exp_abnormal_report_2}.
\item[C2] \textbf{Specialized Terminology}: They also contain specialized networking terminologies that demand domain-specific knowledge for accurate interpretation. For example, prefix notations such as ``\texttt{50.92.0.0/17}'' refer to specific IP address blocks, while AS path entries like ``\texttt{ASpath: 1299 7219 10990}'' trace the sequence of traversed ASes. A robust parsing system must be capable of recognizing semantic variations, such as ``\texttt{AS-path}'' versus ``\texttt{AS PATH}'' as equivalent.
\item[C3] \textbf{Information Noise}: An additional layer of complexity arises from the common practice of email forwarding within these reports as shown in Fig.~\ref{fig:append_detailed_exp_abnormal_report_2}, where embedded message threads and supplementary comments introduce noises that can obscure key event details. This further complicates automated extraction and analysis.
\end{itemize}
The fundamental solution to these challenges lies in accurately interpreting the semantic meaning of each report, including both raw log outputs and domain-specific terminology, much like human network operators do. 
This semantic understanding is crucial for both comprehending the anomalous events and extracting relevant routing information. 
In the era of LLMs such as ChatGPT~\cite{chatgpt}, we now have access to AI systems~\cite{chatgpt} that demonstrate human-like comprehension capabilities and extensive technical knowledge, making them particularly well-suited for automated collection and analysis of routing anomaly events from mailing lists.

\subsubsection{Statistics of Emails}\label{sec:appendix_email_stat}
In Table~\ref{tab:appendix_email_stat}, we summarize the number of emails downloaded from different mailing lists.

\begin{table}[t]
\centering
\caption{The number of emails downloaded from different mailing lists.}
\label{tab:appendix_email_stat}
\scalebox{0.98}{
\begin{tabular}{c||c|c|c}
\hline\hline
No. & Mailing Lists & \begin{tabular}[c]{@{}c@{}}Earliest \\ Recorded Time\end{tabular} & No. of Emails \\ \hline
1   & NANOG         & Jan, 1992               &  286,292                                                                         \\ \hline
2   & RIPE NCC      & Aug., 2003              &  6,831                                                                             \\ \hline
3   & SANOG         & Feb., 2012              &  2,021                                                                                        \\ \hline
4   & AusNOG        & Feb., 2005              &  47,192                                                                                      \\ \hline
5   & AFNOG         & Nov., 2011              &  4,932                                                                                      \\ \hline
6   & INNOG         & July, 2019              &  562                                                                                        \\ \hline
7   & PACNOG        & Aug., 2004              &  2,719                                                                                      \\ \hline
8   & LACNOG        & Dec., 2007              &  10,364                                                                                      \\ \hline
9   & ITNOG         & Mar., 2009              &  1,763                                                                                          \\ \hline
10  & SAFNOG        & Jan., 2014              &  886                                                                                            \\ \hline
11  & SGNOG         & Aug., 2011              &  138                                                                                             \\ \hline
12  & NZNOG         & Feb., 1998              &  10,829                                                                                           \\ \hline
13  & CaribNOG      & Dec., 2017              &  27                                                                                                \\ \hline
14  & Total         & -                       &  374,558                                                                              \\ \hline \hline
\end{tabular}
}
\end{table}

\subsubsection{LLM Prompt for Anomaly Identification and Incident Extraction}
\label{sec:appendix_prompts}

We summarize the prompt and output schema used in our LLM-based incident extraction pipeline. 
We show the prompt template and output JSON schema in Figs.~\ref{fig:prompt_extract} and~\ref{box:prompt_schema}, respectively.
After content normalization, each operator discussion thread is provided to the LLM. 
The LLM performs a joint identification-and-extraction task: it first determines whether the thread describes one or more BGP routing anomaly events; if so, it extracts structured incident records. 
If no routing anomaly event is identified, the model returns an empty incident list.

We use the same prompt template for all LLM variants. The prompt does not assume that an anomaly exists in the input thread. Instead, it explicitly instructs the model to distinguish routing anomalies from unrelated outages, routine maintenance, traffic engineering, debugging discussions, and general operational conversations.

\begin{figure}[t]
\centering
\begin{tcolorbox}[
    colback=gray!5,
    colframe=gray!45,
    boxsep=1mm,
    left=1mm,
    right=1mm,
    top=1mm,
    bottom=1mm,
    arc=1mm
]

\small

\textbf{Role}: You are an expert in inter-domain routing and BGP security.

\vspace{1mm}

\textbf{Task}: Analyze an operator discussion thread and determine whether it describes one or more BGP routing anomaly events.

\vspace{1mm}

\textbf{Anomaly categories:}
\begin{itemize}[leftmargin=*,nosep]
    \item \textbf{Prefix hijack}: an AS announces a prefix or more-specific prefix that it is not authorized to originate.
    \item \textbf{Path hijack}: an AS manipulates or fabricates the AS path while the origin may appear legitimate.
    \item \textbf{Route leak}: an AS propagates routes in violation of expected routing export policies, such as valley-free policy.
\end{itemize}

\vspace{1mm}

\textbf{Instructions:}
\begin{itemize}[leftmargin=*,nosep]
    \item Do not assume that the thread contains a routing anomaly.
    \item Distinguish routing anomalies from unrelated outages, routine maintenance, traffic engineering, debugging discussions, and general operational questions.
    \item Use only information supported by the thread. If a field is missing or ambiguous, return \texttt{null} and list it in \texttt{missing\_fields}.
    \item If the thread discusses multiple distinct incidents, output one record for each incident.
    \item Preserve original AS numbers, prefixes, AS paths, and timestamps whenever they are provided.
    \item Return only valid JSON and do not include text outside the JSON object.
\end{itemize}

\vspace{1mm}

\textbf{Input}: \texttt{[THREAD\_TEXT]}

\vspace{1mm}

\textbf{Output}: Return a JSON object following the schema in Fig.~\ref{box:prompt_schema}. If no routing anomaly is described, return \texttt{contains\_routing\_anomaly=false} and an empty incident list.

\end{tcolorbox}

\vspace{-3mm}
\caption{Prompt template for routing anomaly extraction from operator discussion threads.}
\vspace{-5mm}
\label{fig:prompt_extract}
\end{figure}

\begin{figure}[t]
\centering
\begin{tcolorbox}[
    colback=gray!5,
    colframe=gray!45,
    fonttitle=\bfseries,
    boxsep=1mm,
    left=1mm,
    right=1mm,
    top=1mm,
    bottom=1mm,
    label={box:prompt_schema}
]
\small
\begin{lstlisting}
{
  "contains_routing_anomaly": true | false,
  "incidents": [
    {
      "anomaly_type": "prefix_hijack | path_hijack | route_leak",
      "attacker_or_leaking_asn": "AS number or null",
      "victim_asn": "AS number or null",
      "affected_prefixes": ["prefix", "..."],
      "start_time": "UTC timestamp or null",
      "end_time": "UTC timestamp or null",
      "reported_duration": "duration or null",
      "observed_as_paths": ["AS path", "..."],
      "textual_evidence": ["quoted evidence span", "..."],
      "missing_fields": ["field name", "..."],
      "confidence": "high | medium | low"
    }
  ]
}
\end{lstlisting}

No anomaly case:
\begin{lstlisting}
{
  "contains_routing_anomaly": false,
  "incidents": []
}
\end{lstlisting}
\end{tcolorbox}
\caption{Output JSON schema.}
\label{box:prompt_schema}
\end{figure}

\parab{Output validation.}
We require the LLM output to follow the JSON schema. Outputs that cannot be parsed as valid JSON or that miss required top-level fields are rejected and regenerated. We further apply regex-based validation to routing-specific fields, including AS numbers, IP prefixes, timestamps, and AS paths. Candidate records with malformed, missing, or ambiguous fields are not directly added to the dataset; instead, they are flagged for expert review.

\parab{Cross-model checking.}
To reduce model-specific extraction errors, we run the same prompt on three LLM variants: GPT-5, Gemini-3-Pro, and Claude-4.5-Haiku. We then compare their extracted incident records. Fields that are consistent across models are retained as candidate values, while inconsistent fields are marked for expert review. We do not use an additional LLM to reconcile disagreements; final decisions are made during the expert validation phase.

\parab{Decoding settings.}
We use low-temperature decoding to reduce output variance and improve extraction stability. The settings used in our extraction pipeline are summarized in Table~\ref{tab:llm_prompt_settings}.

\begin{table}[t]
\centering
\caption{LLM settings used for structured incident extraction.}
\label{tab:llm_prompt_settings}
\scalebox{0.87}{
\begin{tabular}{c|c}
\hline\hline
Item & Setting \\ \hline
Input unit & Reconstructed discussion thread \\ \hline
Task & Joint anomaly identification and structured extraction \\ \hline
Output format & JSON \\ \hline
Temperature & 0.3 \\ \hline
Malformed output handling & Regenerate and re-validate \\ \hline
Field validation & JSON schema and regex checks \\ \hline
Models & GPT-5, Gemini-3-Pro, Claude-4.5-Haiku \\ \hline
Disagreement handling & Flagged for expert review \\ \hline \hline
\end{tabular}
}
\end{table}

\subsubsection{Statistics of Collected Anomaly Incidents}\label{sec:ana_collected_data}

\begin{table}[t]
\centering
\caption{The statistics on the types of routing anomaly events we collect.}
\label{tab:stat_anomaly_events}
\scalebox{1.00}{
\begin{tabular}{c||c|c}
\hline\hline
Anomaly Type     & No. of Events & Percentage (\%) \\ \hline
Prefix Hijacking  &  150  &     64.66\%          \\ \hline
Path Hijacking       &  39   &      16.81\%        \\ \hline  
Route Leak       &  43   &    18.53\%          \\ \hline \hline                               
\end{tabular}
}
\end{table}

Here, we present a summary of the statistics for routing anomaly events collected from thirteen public mailing lists. 
Table~\ref{tab:stat_anomaly_events} summarizes the statistics on the types of routing anomaly events.
We can see, for the 232 routing anomaly events, prefix hijacking is the most prevalent, accounting for 64.66\% of the events (150 cases), followed by route leaks at 18.53\% (43 cases), and path hijacking at 16.81\% (39 cases). 

\subsubsection{BGP Data Collection}\label{sec:bgp_evidence}

For each validated incident, we collect the corresponding BGP data from public route collectors. Specifically, we use the reported event time as the reference point and download BGP update data from RouteViews~\cite{routeviews} and RIPE RIS~\cite{ris}. 
Following prior work~\cite{beam2024}, we collect updates within a twelve-hour window before and after each incident to ensure that the abnormal announcements and surrounding routing context are included.
We then filter the raw BGP updates using the validated incident fields, including affected prefixes, attacker or leaking ASNs, victim ASNs, and observed AS paths when available. This step links each incident-level label to the BGP messages that expose the corresponding routing anomaly. It also provides the basis for both event-level and message-level evaluation (\cref{sec:eval}).
We retain the mapping between each incident and its corresponding BGP messages. 
Using the above method, in total, we obtain 164,836 BGP update files with 15,733,920,062 messages, amounting to approximately 1860 GB of data. 

\subsection{Detailed Analysis of General-purpose LLMs' Performance}
\label{app:model_scale_analysis}

To further investigate the non-monotonic relationship between model scale
and routing anomaly detection performance observed in~\cref{sec:llm_perf}, we
analyze four representative general-purpose LLMs:
DeepSeek-R1-Distill-Qwen-7B, Llama-3.1-8B, QwQ-32B, and DeepSeek-R1.
We examine their event-level performance for individual anomaly types,
as well as their output-format compliance and invalid-answer rates.

We define \emph{format compliance} as the fraction of responses that
strictly follow the required output format. The \emph{invalid-answer rate}
is the fraction of responses that cannot be mapped to a valid anomaly
prediction because of malformed, missing, or ambiguous outputs.
Therefore, a response that does not strictly follow the required format
is not necessarily invalid if its prediction can still be unambiguously
identified.

\begin{table}[t]
\centering
\caption{Detailed analysis of representative general-purpose LLMs.
``Compliance'' denotes output-format compliance, and ``Invalid'' denotes the
invalid-answer rate (PH: Prefix Hijacking, PathH: Path Hijacking, RL: Route Leak).}
\label{tab:model_scale_analysis}
\scalebox{0.90}{
\begin{tabular}{c|ccc|c|c}
\hline \hline
\multirow{2}{*}{Models} & \multicolumn{3}{c|}{Event-level F1 by Anomaly Type}                                      & \multirow{2}{*}{Compliance} & \multirow{2}{*}{Invalid} \\ \cline{2-4}
                        & \multicolumn{1}{c|}{PH} & \multicolumn{1}{c|}{PathH} & RL &                             &                          \\ \hline
DS-R1-Qwen-7B           & \multicolumn{1}{c|}{0.5811}           & \multicolumn{1}{c|}{0.4502}         & 0.4917     & 0.9700                      & 0.0235                   \\ \hline
Llama-3.1-8B            & \multicolumn{1}{c|}{0.6012}           & \multicolumn{1}{c|}{0.4698}         & 0.5102     & 0.9795                      & 0.0100                   \\ \hline
QwQ-32B                 & \multicolumn{1}{c|}{0.4200}           & \multicolumn{1}{c|}{0.3105}         & 0.3418     & 0.8573                      & 0.1124                   \\ \hline
DeepSeek-R1             & \multicolumn{1}{c|}{0.6110}           & \multicolumn{1}{c|}{0.4507}         & 0.4983     & 0.9089                      & 0.0617                   \\ \hline \hline
\end{tabular}
}
\end{table}

Table~\ref{tab:model_scale_analysis} provides several additional
observations. First, no model consistently dominates across all anomaly
categories. DeepSeek-R1 achieves the highest F1 score for prefix hijacking,
whereas Llama-3.1-8B achieves the highest F1 scores for path hijacking and
route leak. In contrast, QwQ-32B performs worse than both 7B/8B models
across all three anomaly categories.
Second, the smaller instruction-tuned and distilled models show better
adherence to the required output format. Llama-3.1-8B and
DS-R1-Qwen-7B achieve format-compliance rates of 97.95\% and 97.00\%,
respectively, compared with 85.73\% for QwQ-32B and 90.89\% for
DeepSeek-R1. Their invalid-answer rates are also lower: 1.00\% and
2.35\%, compared with 11.24\% and 6.17\% for QwQ-32B and DeepSeek-R1,
respectively.

Together with the overall accuracy results in Table~II, these results
confirm that increasing model scale does not consistently improve routing
anomaly detection. Model family, post-training, and adherence to the
structured prediction format also affect performance on this task.

\subsection{Extended \sys Design}\label{sec:append_sys_design}
\subsubsection{Training Data Preparation for Tokenizer Training}\label{sec:append_data_token}
To train the tokenizer, we prepare the training data from the collected abnormal BGP data and normal ones introduced in~\cref{sec:eval}, as well as textual anomaly detection instructions.
We use these data including BGP data and texts to make LLMs understand the BGP semantics, while maintaining the general text understanding capacity.
For the anomaly detection instructions, we write 200 instructions with the same semantic meaning in different expression forms.
As for the BGP data, we utilize \texttt{bgpdump}~\cite{bgpdump} to parse the raw data and extract the information within all fields.
We compose a piece of training data using an instruction and a BGP announcement data, as well as the detection output, and creating 22800 in total.
Specifically, to indicate the beginning of the BGP announcement data,
we define an indicator token $<$BGP Update:$>$ in the instruction context, and
each BGP announcement data begins with this indicator.
Similar to the usage of tokenizer, which helps LLMs to learn the semantics of input prompts, we employ instruction learning approach~\cite{instructionlearning2024} to train the tokenizer.
Thus, we create the training data as instruction-output pairs.

\subsubsection{\sys's Fine-tuning Data}\label{sec:sys_tune_data}

We introduce the AS relationship and BGP data for fine-tuning \sys below.

\parab{AS relationship data preparation.}
We utilize the AS relationship dataset provided by CAIDA~\cite{caida-asrelation}, which is updated monthly to capture the dynamic nature of Internet topology. Each entry in the dataset follows the format $<$AS1$|$AS2$|$relationship$>$, where AS1 and AS2 are AS numbers, and the relationship is either customer-to-provider (c2p) or peer-to-peer (p2p). 
To ensure compatibility with LLMs, we convert each raw entry into a natural language sentence. For example, $<$AS1$|$AS2$|$c2p$>$ is transformed into ``AS1 is a customer AS of AS2''. 
This approach enables the model to learn relational knowledge in a human-readable context. 

\parab{BGP data preparation.}
To fine-tune LLMs for routing anomaly detection, we construct a labeled dataset from real-world BGP anomaly events. 
Specifically, for each randomized event-level split, we use 170 of the 249 anomaly
events for training and the remaining 79 events for testing. All abnormal
BGP updates associated with an event remain in the same split.
Among the 249 events, 232 are from our collected dataset (\cref{sec:dataset}), while the remaining 17 are derived from the 18 incidents used in BEAM, with one overlapping event.

To reflect the class imbalance in BGP routing data, we sample normal messages for each anomaly event at an average normal-to-abnormal ratio of 1000:1. Normal updates are drawn from the same temporal windows and related AS paths as the corresponding abnormal updates to preserve contextual similarity while maintaining benign behavior. The 1000:1 ratio refers to the message ratio within each anomaly event. To avoid data leakage, all abnormal messages associated with the same event are kept within a single data split, and the normal messages associated with the event remain in the corresponding split.

For each BGP update, we construct an instruction-following training instance by pairing the update with a routing anomaly detection prompt. The prompt asks the model to determine whether the update is anomalous and, if so, identify its anomaly type. Each instance is represented as a triplet $(I_i, U_i, Y_i)$, where $I_i$ denotes the detection instruction, $U_i$ the BGP update, and $Y_i$ the ground-truth label (\eg normal, prefix hijacking, path hijacking, or route leak). These instruction-tuning data enable the LLM to learn routing anomaly detection directly from BGP update semantics.

\subsubsection{Details of Routing-domain Adaptation}
\label{app:routing_adaptation}
We provide the details of the two-stage fine-tuning procedure below.

\parab{Stage 1: AS relationship fine-tuning.}
We collect the latest CAIDA AS relationship data and construct sentences describing each relationship (e.g., ``AS1 is a provider AS of AS2''). During training, we randomly masked 15\% of tokens in each input sentence; of the chosen tokens, 85\% were replaced with \texttt{[MASK]}, while 15\% were left unchanged. 
We train the model to reconstruct the original tokens at the selected positions using the negative log-likelihood loss:
\begin{equation}
    \mathcal{L}_{\mathrm{mask}} = -\sum_{i\in\mathcal{M}}\log P(x_i\mid\hat{X};\theta), \nonumber
\end{equation}
where $\mathcal{M}$ denotes the selected token positions, $\hat{X}$ is the masked input, and $x_i$ is the original token at position $i$.

\parab{Stage 2: BGP data fine-tuning.}
We use the BGP dataset prepared as described in~\cref{sec:sys_tune_data}. The model from stage 1 was fine-tuned with LoRA~\cite{lora2021} applied
to the attention projection layers and MLP projection layers. The cross-entropy loss for multi-class anomaly classification was minimized:
\begin{equation}
    \mathcal{L}_{CE} = -\frac{1}{N}\sum_{j=1}^N\sum_{i=1}^C y_{j,i}\log(p_{j,i}),  \nonumber
\end{equation}
where $C$ is the number of anomaly types, $y_{j,i}$ is the true label, and $p_{j,i}$ the predicted probability.

\subsubsection{Details of Time-aware Routing Knowledge Retrieval}
\label{app:rag}

This section provides implementation details for the time‑aware routing knowledge retriever.

\parab{Knowledge base construction.}
The external knowledge base $\kappa$ is built from three primary sources:
\begin{icompact}
    \item \textbf{AS relationships.} We use monthly CAIDA AS relationship snapshots~\cite{caida-asrelation}. Each relationship tuple $\langle \text{AS}_1, \text{AS}_2, \text{rel} \rangle$ is converted into a natural language sentence using templates such as ``\texttt{AS1 is a customer of AS2}'' for provider‑customer links, ``\texttt{AS1 and AS2 are peers}'' for peer‑peer links, and ``\texttt{AS1 and AS2 are siblings}'' for sibling links. These sentences mirror the format used during routing‑domain adaptation (\cref{sec:fine_tune_design}) to maintain consistency.
    \item \textbf{Prefix‑origin histories.} From historical BGP RIBs and updates (RouteViews and RIS), we extract every distinct prefix‑origin AS pair observed per month. For each prefix, we record the set of ASes that originated it and the last seen timestamp. We then generate evidence sentences such as ``\texttt{AS15169 typically announces prefix 2001:4860::/32}'' if the AS was a regular origin for that prefix over the preceding months. Regularity is determined by a simple frequency threshold: an origin is considered ``typical'' if it announced the prefix in at least 80\% of the monthly snapshots within the one‑year window before the event.
    \item \textbf{AS‑path edges and RPKI/IRR data.} We extract all adjacent AS pairs observed in AS paths during the same historical windows, building a set of previously seen AS‑level edges. For RPKI, we parse ROA files archived by public repositories (e.g., RIPE NCC) to record which prefix‑origin pairs were covered by a valid ROA at each point in time. IRR records are similarly timestamped by the date of the last update.
\end{icompact}
All evidence records are associated with a validity interval or an observation timestamp. This ensures that retrieval can later filter evidence to match the event time.

\parab{Indexing and retrieval.}
We encode every evidence record into a dense vector using a pre‑trained Sentence Transformer model. The encoding is purely text‑based: each evidence sentence (e.g., ``\texttt{AS1299 is a customer of AS7219}'' or ``\texttt{AS15169 typically announces prefix 2001:4860::/32}'') is passed through the model to obtain a fixed‑size embedding.

All embeddings are stored in a FAISS index with exact nearest‑neighbour search~\cite{faiss2024}. During inference, given a BGP update $m$, we construct a query string that concatenates the prefix, origin AS, and the full AS path. This query is encoded with the same Sentence Transformer. We then retrieve the top‑$K$ most similar evidence records according to cosine similarity:

\begin{equation}
    \mathcal{R}(x) = \text{TopK}_{e_i \in \kappa} \left( \text{sim}(x, e_i) \right), \nonumber
\end{equation}
where $x$ is the embedding of the query, $e_i$ are the embeddings of the evidence entries, $\texttt{sim}(\cdot,\cdot)$ denotes cosine similarity, and $K$ is a constant hyperparameter. We further filter the retrieved candidates to keep only those whose validity interval contains the timestamp of $m$, ensuring time consistency.

\parab{Parameter settings.}
We set the retrieval depth $K = 5$ in all experiments. This value was chosen empirically as a trade‑off between providing sufficient context and avoiding prompt length explosion. The knowledge base is rebuilt on a monthly basis to incorporate new CAIDA relationship snapshots and the latest historical BGP archives. The FAISS index is regenerated after each rebuild. Sentence Transformer embeddings are kept fixed and are not fine‑tuned.

\begin{table*}[t]
\centering
\caption{95\% confidence intervals of the event-level and message-level
routing anomaly detection performance reported in
Table~\ref{tab:overall_performance}.}
\label{tab:overall_performance_ci}

\setlength{\tabcolsep}{4pt}
\renewcommand{\arraystretch}{1.05}

\begin{tabular}{c|c||c|c|c|c|c}
\hline\hline
\multicolumn{2}{c||}{{Metric}} &
{\sys} & {BEAM} &
{MS-LSTM} & {RF} & {GCN} \\
\hline

\multirow{6}{*}{{Event-level}}
& AC
& {[0.8566, 0.8860]}
& [0.5366, 0.5856]
& [0.3533, 0.4061]
& [0.4647, 0.5099]
& [0.4919, 0.5329] \\ \cline{2-7}

& PR
& {[0.8242, 0.8598]}
& [0.5270, 0.5794]
& [0.3022, 0.3598]
& [0.4434, 0.4926]
& [0.4662, 0.5112] \\ \cline{2-7}

& RC
& {[0.8634, 0.8956]}
& [0.5674, 0.6174]
& [0.3294, 0.3842]
& [0.4777, 0.5247]
& [0.5041, 0.5471] \\ \cline{2-7}

& F1
& {[0.8400, 0.8742]}
& [0.5363, 0.5875]
& [0.3023, 0.3571]
& [0.4515, 0.4987]
& [0.4824, 0.5246] \\ \cline{2-7}

& FPR
& {[0.0629, 0.0841]}
& [0.1948, 0.2320]
& [0.3062, 0.3512]
& [0.2409, 0.2761]
& [0.1750, 0.2076] \\ \cline{2-7}

& FNR
& {[0.1044, 0.1366]}
& [0.3826, 0.4326]
& [0.6158, 0.6706]
& [0.4753, 0.5223]
& [0.4529, 0.4959] \\
\hline

\multirow{6}{*}{{Message-level}}
& AC
& {[0.9354, 0.9576]}
& [0.4100, 0.4426]
& [0.4947, 0.5391]
& [0.4678, 0.5046]
& [0.5443, 0.5793] \\ \cline{2-7}

& PR
& {[0.8833, 0.9119]}
& [0.2308, 0.2754]
& [0.4144, 0.4648]
& [0.3975, 0.4399]
& [0.4681, 0.5071] \\ \cline{2-7}

& RC
& {[0.9093, 0.9331]}
& [0.1797, 0.2239]
& [0.3356, 0.3826]
& [0.3819, 0.4211]
& [0.4506, 0.4890] \\ \cline{2-7}

& F1
& {[0.8870, 0.9124]}
& [0.1843, 0.2313]
& [0.3444, 0.3938]
& [0.3796, 0.4196]
& [0.4506, 0.4896] \\ \cline{2-7}

& FPR
& {[0.0162, 0.0268]}
& [0.2327, 0.2647]
& [0.1730, 0.2098]
& [0.1621, 0.1917]
& [0.1354, 0.1610] \\ \cline{2-7}

& FNR
& {[0.0669, 0.0907]}
& [0.7761, 0.8203]
& [0.6174, 0.6644]
& [0.5789, 0.6181]
& [0.5110, 0.5494] \\
\hline\hline
\end{tabular}

\end{table*}

\subsubsection{Evidence-grounded Anomaly Reasoning}
\label{app:evidence_reasoning}

After constructing the structured BGP representation and retrieving routing evidence,
\sys performs anomaly detection using the routing-adapted LLM. The input prompt contains
three parts: a task instruction, the structured BGP update, and the retrieved evidence
block. The model is asked to output both an anomaly label and the evidence supporting
the decision.
The output is constrained to a structured format, as shown in Fig.~\ref{fig:output_format}.

\begin{figure}
\centering
\begin{lstlisting}
{
  "is_anomaly": true,
  "anomaly_type": "prefix_hijack",
  "confidence": 0.91,
  "evidence": [
    "Observed origin AS10990 differs from historical origin",
    "Affected prefix 50.92.0.0/17 was reported by operators",
    "The anomalous update appears during the incident window"
  ]
}
\end{lstlisting}
\label{fig:evidence_reasoning_output}
\caption{Structured evidence-grounded detection output.}
\label{fig:output_format}
\end{figure}

This evidence-grounded output serves two purposes. First, it improves robustness by
encouraging the model to base its prediction on retrieved routing facts rather than
memorized patterns or unsupported textual associations. Second, it improves operational
usefulness because network operators need to understand why an alert is raised before
taking mitigation actions.

\subsubsection{Discussion of Design Choices}\label{sec:design_discussion}
We discuss the design choices of \sys in the following.

\parab{Why not use rules alone?}
Rule-based detectors can capture specific symptoms, such as RPKI-invalid origins or valley-free violations. However, real-world incidents often involve incomplete, noisy, or conflicting evidence. For example, a route leak may not violate a single deterministic rule under incomplete AS relationship data, and a prefix hijack may resemble a legitimate MOAS event without historical context. \sys uses LLMs to integrate heterogeneous evidence, while still grounding the reasoning process in structured routing facts.

\parab{Why not use vanilla LLMs directly?}
Vanilla LLMs have broad language understanding, but BGP anomaly detection requires precise interpretation of protocol fields and dynamic routing evidence. Our preliminary study shows that vanilla LLMs can detect some events but are unstable across models and messages. \sys addresses this by explicitly representing BGP-specific structure, adapting the model to routing domain, and retrieving time-consistent evidence during inference.

\parab{Why use time-aware retrieval?}
Internet routing conditions change over time. A prefix-origin pair that is suspicious in one year may become legitimate later, and AS relationships may change across snapshots. Time-aware retrieval ensures that \sys evaluates each update using evidence that is consistent with the update time, which is essential for both offline benchmark evaluation and realistic deployment.

\subsection{Extended System Implementation}\label{sec:more_imp}

\parab{Routing anomaly incident data collection.}
To gather emails from public mailing lists, we use \texttt{BeautifulSoup} to implement an email crawler ($\sim$204 LOC) that automatically collects emails from thirteen public lists excluding RIPE NCC. Due to RIPE NCC’s unique structure, we develop a specialized crawler for it. 
In addition, due to the structural changes and the adoption of HyperKitty for web archiving in the NANOG Mailing List Archive starting in 2025, we develop a dedicated scraping script to accommodate these updates.
Emails are stored in CSV format, recording key fields such as \texttt{email\_id}, \texttt{thread\_id}, \texttt{level}, \texttt{time}, \texttt{email\_subject}, \texttt{author}, \texttt{email\_address}, and \texttt{email\_content}.

To extract routing anomaly events, we use an additional script ($\sim$186 LOC) to invoke LLMs' APIs for deeper semantic analysis. To enrich incomplete anomaly events, we retrieve missing data from the IRR and RPKI database based on other entries within the same event, implemented in $\sim$127 LOC.

\parab{Analysis of LLM outputs.}
To evaluate routing anomaly detection performance and standardize the results, we develop a dedicated testing script that feeds test data to LLMs via API calls and enforces a structured JSON output format. 
During testing, we observe that LLMs often generate semantically correct but lexically inconsistent outputs. For instance, when prompted to return ``Prefix Hijacking'' for a specific anomaly, the model may instead output variations such as ``Prefix Hijack'', ``BGP Hijack'', or ``prefix\_hijacking''.
To account for this, we measure detection performance based on semantic accuracy rather than strict string matching. 
Notably, the fine-tuned model shows significantly improved consistency, adhering more closely to the expected output format specified in the prompt.

\subsection{Extended Evaluation}\label{sec:appendix_eval}
To further quantify the variability of the evaluation results, we report
the 95\% confidence intervals (CIs) of the metrics in
Table~\ref{tab:overall_performance}. For each metric, we first compute its
value independently from the predictions of each of the five randomized
event-level train/test splits. We then compute the confidence interval of
the mean across the five splits using the Student's $t$-distribution:

\begin{equation}
    \bar{x} \pm t_{0.975,n-1}\frac{s}{\sqrt{n}}, \nonumber
\end{equation}

where $\bar{x}$ and $s$ denote the sample mean and standard deviation across
the splits, respectively, and $n=5$. For a two-sided 95\% confidence
interval, the corresponding critical value with four degrees of freedom is
$t_{0.975,4}=2.776$~\cite{nist_t_distribution}. Table~\ref{tab:overall_performance_ci}
reports the resulting confidence intervals.

\begin{figure*}[t]
\centering
\begin{subfigure}{1\textwidth}
\centering
\begin{lstlisting}
(*\bfseries Subject*): BGP route hijack by AS10990
(*\bfseries From*): Clinton Work clinton at scripty.com
(*\bfseries Time*): (*\textit{Thu Jul 30 02:46:55 UTC 2020}*)
(*\bfseries Contents*): We saw a bunch of our IP blocks hijacked by AS10990 from 19:15 MDT until 20:23 MDT. Anybody else have problems with that.

ASpath:  1299 7219 10990

50.92.0.0/17	AS10990
198.166.0.0/17	 AS10990
198.166.128.0/17	AS10990
162.157.128.0/17	AS10990
162.157.0.0/17	AS10990
50.92.128.0/17	AS10990

--
Clinton Work
Airdrie, AB

(*\bfseries Subject*): Re: BGP route hijack by AS10990
(*\bfseries From*): Jeff Bilyk jbilyk at gmail.com
(*\bfseries Time*): (*\textit{Thu Jul 30 04:26:21 UTC 2020}*)
(*\bfseries Contents*): We appeared to be impacted with some address space within 206.47.0.0/16
which AS577 normally advertises, but that was between 15:50 and 16:30
Eastern.

Jeff
\end{lstlisting}
\vspace{-3mm}
\caption{Prefix hijacking~\cite{hijackexp}}
\label{fig:app_exp_bgphijacking}
\end{subfigure}
\begin{subfigure}{1\textwidth}
\centering
\begin{lstlisting}
(*\bfseries Subject*): Route Leak? AS11845 / Vox Telecom Ltd
(*\bfseries From*): Phil Lavin phil.lavin at vonage.com
(*\bfseries Time*): (*\textit{Wed Jun 21 13:13:58 UTC 2023}*)
(*\bfseries Contents*): Hi Folks, Seeing traffic from AWS use1 (216.147.2.235) to UK (62.3.100.19) has been attempting to transit via AS11845 for the last few hours. Looks like a route leak from Vox->AWS at DE-CIX. 

traceroute to 62.3.100.19 (62.3.100.19), 30 hops max, 60 byte packets
1  ec2-3-236-61-157.compute-1.amazonaws.com (3.236.61.157)  2.505 ms  2.001 ms  1.988 ms
2  240.0.228.64 (240.0.228.64)  0.348 ms  0.443 ms  0.440 ms
3  240.0.228.89 (240.0.228.89)  0.334 ms  0.331 ms  0.430 ms
4  240.192.54.146 (240.192.54.146)  0.260 ms  0.310 ms  0.307 ms
5  240.0.228.57 (240.0.228.57)  0.250 ms  0.309 ms  0.305 ms
6  240.0.228.34 (240.0.228.34)  0.291 ms  0.298 ms  0.289 ms
7  240.0.48.14 (240.0.48.14)  0.832 ms  0.829 ms  0.826 ms
8  240.0.48.19 (240.0.48.19)  0.307 ms  0.400 ms  0.392 ms
9  240.192.15.51 (240.192.15.51)  0.343 ms  0.419 ms  0.456 ms
...
18  voxtelecom.co.za (206.82.105.31)  229.508 ms  229.811 ms  229.798 ms
19  41-193-120-5.vox.co.za (41.193.120.5)  228.144 ms  228.141 ms *
20  * * *
21  linx-2.zen.net.uk (195.66.236.158)  228.036 ms  228.154 ms *
22  * lag-6.p2.thn-lon.zen.net.uk (51.148.73.166)  223.587 ms *
23  * * *
27  * 62-3-100-19.dsl.in-addr.zen.co.uk (62.3.100.19)  230.499 ms  230.708 ms

Can anybody see the extent of the leak in your route tables? As it stands, my home Internet is unbearable.

Phil


(*\bfseries Subject*): Re: Route Leak? AS11845 / Vox Telecom Ltd
(*\bfseries From*): Phil Lavin phil.lavin at vonage.com
(*\bfseries Time*): (*\textit{Wed Jun 21 14:13:35 UTC 2023}*)
(*\bfseries Contents*): This seems to have been resolved and stable for the past 30 mins.

Phil
\end{lstlisting}
\vspace{-3mm}
\caption{Route leak~\cite{routeexp}}
\label{fig:app_exp_leak}
\end{subfigure}
\vspace{-5mm}
\caption{More details about the examples for BGP prefix hijacking and route leak reports from NANOG mailing list~\cite{hijackexp,routeexp}.}
\label{fig:append_detailed_exp_abnormal_report_1}
\end{figure*}

\begin{figure*}[t]
\centering
\begin{lstlisting}
(*\bfseries Subject*): BGP Hijack/Sickness with AS4637
(*\bfseries From*): Brad Hooper brad at coloau.com.au
(*\bfseries Time*): (*\textit{Fri Jun 1 02:56:22 UTC 2018}*)
(*\bfseries Contents*): Hello,

Just to clear up a few things. We are not running any route optimization software (ever). The reason we "refused" to help is because we were not going to contact our transit providers NOC regarding other parties routes, even if we did they wouldn't be of assistance.

We are purely passing on the routes we receive from our transit providers to our customers. We are not modifying the routes in any way shape or form.

We incest routes from a lot of transit providers and send most of the data to route views (As do a number of our customers) which is why this route was seen from us.

I have completed a soft clear on our BGP session towards AS4637 and the route still exists. Sorry we can't be of assistance in this case but this is fully out of our control.

xxxx at re0-cr1.ty8.ty.jp> show route 128.10.4.0/24 detail

vrf-international.inet.0: 696465 destinations, 1194388 routes (696461 
active, 0 holddown, 4 hidden)
128.10.4.0/24 (1 entry, 1 announced)
         *BGP    Preference: 170/-101
                 Next hop type: Router, Next hop index: 790
                 Address: 0xff29810
                 Next-hop reference count: 1279932
                 Source: 202.127.69.33
                 Next hop: 202.127.69.33 via ae0.401, selected
                 Session Id: 0x181
                 State: <Active Ext>
                 Peer AS:  4637
                 Age: 2w4d 11:08:17
                 Validation State: unverified
                 Task: BGP_4637.202.127.69.33
                 Announcement bits (4): 1-KRT 2-BGP Route Target 
5-BGP_RT_Background 6-Resolve tree 6
                 AS path: 4637 3257 29909 16532 16532 16532 16532 I
                 Communities: 4637:32031 4637:32314 4637:32504 4637:60952
                 Accepted
                 Localpref: 100
                 Router ID: 202.84.219.12

Regards,
Brad
ColoAU (AS63956)

Colocation Australia Pty Ltd
On 01/06/18 06:36, Job Snijders wrote:
> On Thu, May 31, 2018 at 02:40:06PM +0000, Job Snijders wrote:
>> 
>> Upon further inspection, it seems more likely that the bgp optimiser is in ColoAU's network. Given the scale of AS 4637, if it were deployed inside Telstra I'd expect more problem reports. AS 4637 may actually just be an innocent bystander.
>> It is interesting to note that the /23 only appears on their Sydney based routers on https://lg.coloau.com.au/
>> Is ColoAU's refusal to cooperate a matter of misunderstanding? Perhaps you should just straight up ask whether they use any type of "network optimisation" appliance. I found a few more interesting routes inside ColoAU's looking glass:
>> 128.10.4.0/24 - AS_PATH 63956 4637 3257 29909 16532 16532 16532 16532 (should be 128.10.0.0/16 originated by AS 17, Purdue University)
>> 192.54.130.0/24 - AS path: 135069 9439 (does not exist in the DFZ, a peering lan prefix? a typo?)
>> 67.215.73.0/24 - AS path: 2764 1221 36692 (does not exist in the DFZ, a peering lan prefix? a typo?)
>> ColoAU propagated the above routes to their transit customers, so the 128.10.4.0/24 and 18.29.238.0/23 announcements definitely count as BGP hijacks with fabricated an AS_PATH.
>>
>> Kind regards,
>> Job

\end{lstlisting}
\vspace{-3mm}
\caption{An example of BGP path hijacking event report from NANOG mailing list~\cite{bgphijack_2018}.}
\label{fig:append_detailed_exp_abnormal_report_2}
\end{figure*}

\end{document}